\newcommand{\beq}{\begin{equation}}
\newcommand{\eeq}{\end{equation}}
\newcommand{\bea}{\begin{eqnarray}}
\newcommand{\eea}{\end{eqnarray}}
\newcommand{\bear}{\begin{eqnarray*}}
\newcommand{\eear}{\end{eqnarray*}}

\newcommand{\rf}[1]{(\ref{#1})}

\def\bra#1{\langle #1|}
\def\ket#1{|#1\rangle}
\def\sstyle{\scriptstyle}

\documentclass[12pt]{iopart}
\usepackage{epsfig}

\usepackage{psfrag}
\usepackage{graphicx}

\begin{document}
%\documentstyle[prl,aps,twocolumn]{revtex}
%\documentstyle[aps,multicol]{revtex}
%\documentstyle [12pt,titlepage]{article} 
%\draft
\title
{Different facets of the raise and peel model}
\author{Francisco  C. Alcaraz$^1$ and Vladimir Rittenberg$^{2,3}$}
\address{$^1$Instituto de F\'{\i}sica de S\~ao Carlos, Universidade de S\~ao 
Paulo,
Caixa Postal 369, 13560-590, S\~ao Carlos, SP, Brazil}
\ead{alcaraz@if.sc.usp.br}
\address{$^2$Physikalisches Institut, Bonn University, 53115 Bonn, Germany}
\address{$^3$ Department of Mathematics and Statistics, University
of Melbourne, Parkville,   Victoria 3010, Australia}
\ead{vladimir@th.physik.uni-bonn.de}

\begin{abstract}

The raise and peel model is a one-dimensional stochastic model of a 
fluctuating interface with nonlocal interactions. This is an interesting 
physical model, 
in this paper we review its properties. 
It's phase diagram  has a massive phase 
and a gapless phase with varying critical exponents. At the phase 
transition point, the model exhibits conformal invariance which is a  
space-time symmetry. Also at this point the model has several other facets 
which are the connections to associative algebras, two-dimensional fully 
packed loop models and combinatorics.    
\end{abstract}


\date{\today}
\maketitle

%\begin{multicols}{2}
%\narrowtext    
\section{ Introduction}
The raise and peel model (RPM), 
reviewed in this paper, is a one-dimensional 
adsorption-desorption model of a fluctuating 
interface \cite{GNPR1,ALR}. The interface 
evolves following nonlocal Markovian dynamics. Originally, the model "appeared" 
\cite{GNPR2} during an investigation of some intriguing connections between the 
groundstate wavefunctions of the XXZ quantum chain and two-dimensional 
dense O($n=1$) fully packed loop models \cite{BGN,PRGN}. For some special 
boundary fully packed loop (FPL) models \cite{EKLP,BBNY} are related \cite{DG} to Alternating Sign
Matrices \cite{BR}, which are of interest in combinatorics. These magic 
connections were found by Razumov and Stroganov \cite{RS1}-\cite{RS3}.

 The stationary states probability distribution functions (PDF) of the RPM are
given precisely by the Razumov-Stroganov wavefunctions. It was later
understood that the RPM is in fact a very interesting model on its own.
Moreover, objects with a nice mathematical structure, and this indeed the 
case of Razumov-Stroganov wavefunctions, allow, using small systems sizes, 
to make conjectures for the expression of physical quantities for any 
systems sizes.  

 The RPM is not only special because of the nice properties of the
stationary states PDF but also for its dynamics. For large system sizes, 
the finite-size scaling eigenspectrum of the Hamiltonian describing the 
continuous time evolution of the RPM is expressed in terms of characters 
of the Virasoro algebra \cite{CI}, hence conformal invariance (this implies 
that the dynamical critical exponent has the value $z = 1$). To our knowledge, the RPM is 
the first known example of a stochastic process having this space-time 
symmetry. As we are going to see, this symmetry has consequences for the
physical properties of the model including the stationary states.

 By changing the ratio of the adsorption and desorption rates of the model, 
the RPM can be taken away from the Razumov-Stroganov point (which corresponds
to the case where the rates are equal). We show the phase diagram of the 
model thus obtained. If the adsorption rates exceed the desorption rates, one
gets a gapless phase with continuously varying critical exponents. If the 
desorption rates are higher than the adsorption ones, the system is 
massive (finite correlation lengths).

 The presentation of the different facets of the RPM will be kept as 
elementary as possible, for the reader who wants to know more, we give 
references.

 In Section 2 we give a connection between associative algebras and 
stochastic processes. This connection is important since the structure 
of the RPM and some of its extensions are related to the Temperley-Lieb (TL) algebra 
\cite{TL,PM} and its extensions \cite{DG}. Moreover one can use two representations 
of the TL algebra. One in terms of RSOS paths which is useful for the interface 
model and another one in terms of up and down spins. In the latter 
representation, one gets a quantum spin chain Hamiltonian which is 
integrable and therefore the spectrum can be computed exactly using 
the Bethe ansatz. In 
Section 3 we give the finite-size scaling limit of the Hamiltonian
eigenspectrum. It is at this point that conformal invariance enters
 the 
picture.

 In Section 4 we show how the Hamiltonian expressed in terms of the 
generators of the TL algebra acts in the representation given by link 
patterns (equivalent to RSOS paths). We consider the example of a 
Hamiltonian containing 5 generators acting in a space of 5 
link patterns configurations. We compute the PDF of the stationary state 
(ground-state wavefunction) and show how it can be related to the 26 
configurations of a fully packed loop model (FPL) on a rectangle or to  
26 vertically symmetric alternating sign matrices. This phenomenon which 
is valid not only for the case of five generators, is very 
interesting since it relates the stationary state of a one-dimensional 
stochastic process to an equilibrium two-dimensional system. This is the 
Razumov-Stroganov conjecture.

 All these considerations were valid for the RPM at the Razumov-Stroganov 
point. In Section 5 we describe the one-parameter dependent RPM. The 
parameter denoted by $u$ is the ratio of the adsorption to desorption rates.
For $u$ not equal to one, the model is not anymore related to the TL 
algebra, few analytic calculations are possible and the study is based on 
Monte-Carlo simulations. We give the phase diagram of the model which is 
based on the results given in the next sections.

 In Section 6 we give the properties of the stationary states, stressing 
what is special for the case $u = 1$. Section 7 deals with the dynamics of 
the model. Since the desorption processes are not local, avalanches occur. 
We briefly mention their properties.
 
Some results not described in this paper are briefly presented in Section 
8. We also list some open questions.

\section{ Markovian dynamics and associative algebras.}\label{sect2}

 The continuous time evolution of a system composed by the states 
$a = 1,2,\ldots,N$
with probabilities $P_a(t)$ is given by a master equation that can be interpreted 
as an imaginary time Schr\"odinger equation:
\begin{equation}\label{2.1}
\frac{d}{dt} P_a(t) = -\sum_b H_{a,b} P_b(t),
\end{equation}
where the Hamiltonian $H$ is an $N\times N$ intensity matrix: $H_{a,b}$ 
nonpositive ($a \neq b$)
 and $\sum_a H_{a,b} = 0$. $-H_{a,b}$ is the rate for the transition
$\ket b \rightarrow \ket a$. The ground-state wavefunction of the system $\ket0$, $H \ket0 = 0$, gives
the probabilities in the stationary state:
\begin{equation} \label{2.2}
\ket0 = \sum_a P_a \ket a,\;\;\;\;\;\; P_a = \lim_{t \to \infty}  P_a(t). 
\end{equation}

 The normalization factor of the unnormalized probabilities $P_a$ is
$\bra0 \ket0$ where 
\begin{equation} \label{2.2p}
\bra0 = \sum_a \bra a, \;\;\;\;\;  \bra0 H = 0.
\end{equation}

 Since $H$ is an intensity matrix the real parts of the $N$ eigenvalues $E(k)$  
are nonnegative and, if complex, the eigenvalues come in conjugate pairs. 
The eigenvalue zero is not degenerate. Conversely, if the spectrum of a matrix 
has these properties, through similarity transformations the matrix  can be 
brought to the form of intensity matrices (the solutions are not unique).

 Let us now consider an associative algebra with generators $A(i)$ 
($i=1,2,\ldots,m$). Taking products of the generators one gets the words $w(r)$. 
The algebra is defined by giving some relations between the words. If we 
can choose the independent words $W(r)$ such that any product of them  verify the 
relation:
\begin{equation} \label{2.3}
W(r)W(s) =  \sum_q p_q^{r,s}W(q),\;\;\;\;\;  p_q^{r,s}\geq 0,\;\;\;\;\sum_q
 p_q^{r,s} = 1,
\end{equation}
then the Hamiltonian
\begin{equation} \label{2.4}
H = \sum_u a(u) (1 - W(u)),
\end{equation}
acting in the vector space 
defined by the basis of independent  words $W(s)$,  is an intensity matrix 
if the coefficients $a(u)$ are nonnegative. $H$ is acting from the left on the 
words $W(s)$ of the vector space: $W(u)W(s)$ ($W(u)$ is one on the terms in 
\rf{2.4} and $W(s)$ belongs to the vector space). The action of the independent 
words $W(u)$ on the vector space defined by the same words gives the regular 
representation of the algebra. 
 
If the algebra $A$ contains a 
left ideal\footnote{A left $I_A$ ideal of an algebra $A$ is a subspace of $A$ such that $aX \in I_A$ for all 
$a \in A$ and $X \in I_A$.} 
 defined by the words
 $I_A$, the Hamiltonian \rf{2.4} acting 
on this ideal gives again a stochastic process. If $A$ has several ideals
$I_1 ( I_2 (\cdots(I_n$, $H$ has a block triangular form. The ground-state 
wavefunction $|0>$ is a linear combination of the states (words) which 
define the vector space $I_1$. These states are also called recurrent 
states. The words belonging to the ideals $I_2,\ldots,I_n$ but not belonging to 
$I_1$ do not appear in the stationary state.
 
We will give a simple example. Consider the Abelian algebra given by $m$ 
generators $A(i)$ ($i = 1,2,\ldots,L$) satisfying the relations:
\begin{equation}\label{2.5}
A(i)^2 = a A(i+1)^2 +b A(i)A(i+1), \;\;\; i =1,2,\ldots,L-1,
\end{equation}
\begin{equation}\label{2.6}
A(L)^2 = a + b A(L),       
\end{equation}
where 
\begin{equation}\label{2.7}
[A(i),A(j)] = 0, \;\;\;\; i,j=1,\ldots,L,           
\;\;\;\;\;  a+b =1,\;\; a,b\geq0.
\end{equation}

The model is defined by the Hamiltonian:
\begin{equation} \label{2.8}
H = 1 - \frac{1}{L} \sum_{i=1}^L A(i).    
\end{equation}
$H$ acts in the $2^L$ dimensional vector space of independent words which is 
given by the monomials: $1, A(i), A(i)A(j)$ ($i\neq j$), $\ldots$. 
 The physical
interpretation of the vector space is obvious. Take a one-dimensional system
with $L$ sites. Each site   can be empty or occupied by at most one particle. In the 
state $A(i_1)A(i_2)\cdots A(i_k)$ the site $i_n$  is occupied if $A(i_n)$ appears in 
the expression of the monomial. In this sandpile model (see Dhar's papers 
\cite{DD} for much more on this subject), the toppling mechanism is encoded in 
the algebra \rf{2.5}-\rf{2.7} and the dynamics in Eq. \rf{2.8}.
 In a unit of time a particle is introduced with probability 
$1/L$ at any site of the lattice (see Eq.~\rf{2.8}). Without increasing the 
time, a given site $i$ with more than a single particle, has the probability 
$a$  (probability $b$) of sending two (one) of its particles to the site $i+1$ 
(see Eq.~\rf{2.5}). Particles at the site $L$ may leave the lattice (see 
Eq.~\rf{2.6}). The process continues up to when we have at most a single particle 
in any lattice point.

 A special class of associative algebras in which the condition \rf{2.3} is 
satisfied are semigroup algebras. In this case, the relations \rf{2.3} take 
the simple form:
\begin{equation} \label{2.9}
U(q) = W(s).
\end{equation}
An important example of this kind is the Temperley-Lieb (TL) algebra with the 
generators $e_j$ ($j=1,\ldots,L-1$) satisfying:
\beq\label{2.10}
e_j^2 = (q+q^{-1})e_j,\qquad e_je_{j\pm1}e_j = e_j, \qquad e_je_k = e_ke_j \quad
{\rm for} \quad |j-k| > 1,
\eeq
with $q$ having the special value $q = \exp(i\pi/3)$. The special choice in \rf{2.4} where we take $a(u) =1$ and  for $W(u)$ only the generators $e_j$  gives the exact integrable quantum Hamiltonian

\beq\label{2.11}
H = \sum_{j=1}^{L-1} (1-e_j).
\eeq
For other values of $q$, the relations \rf{2.10} define an associative algebra 
but not a semigroup. The TL semigroup \rf{2.10} can be generalized in several 
ways. One can add boundary generators \cite{MS}-\cite{APR} or 
one can  generalize the group algebra itself. Examples of the latter 
are the multi-colored versions of the TL semigroup \cite{BJ,PDF} or the 
rotor model \cite{BAT}. Physical applications of these models have not yet been 
studied.
 
Another important example of a semigroup with possible applications to 
stochastic processes is the Brauer semigroup  algebra \cite{MR,MNR,dGN}.

 In all these cases, the 
unnormalized 
probabilities $P_a$ which appear in the ground-state 
wavefunction $|0>$ are positive numbers. If these numbers are integer multiples of 
the lowest one, 
 they 
might have a combinatorial meaning and the challenge is to find and 
understand them.
 
From now one we will study only the applications of the TL semigroup 
defined by Eq. \rf{2.10} and take the Hamiltonian \rf{2.11} with $L$ an even 
number ($L=2n$).

 One has to specify the vector space in which the Hamiltonian \rf{2.11}
acts. The TL algebra has a left ideal $I_1$ defined by the words 
\cite{PM,PRGN}

 \begin{equation} \label{2.12}
W(s)J_0, \;\;\;\;\; J_0= \prod_{j=1}^{L/2} e_{2j-1}, 
\end{equation}
where $W(s)$ is any word of the algebra.
There are                 
\begin{equation}\label{2.13}
C_n = \frac{1}{n+1} {2n \choose n}
\end{equation}
words in this ideal. For example, if $L = 4$, the two independent words in 
this left ideal are: $e(1)e(3)$ and $e(2)e(1)e(3)$. In Section 4 we discuss in 
detail the action of $H$ in this vector space. The raise and peel model at the 
Razumov-Stroganov point is defined by the Hamiltonian \rf{2.11} acting on 
this ideal.

 As opposed to the Abelian algebra \rf{2.5}-\rf{2.7} or other Abelian sandpile 
models, it is not clear at all how the TL algebra which is not Abelian can 
be related to a toppling process.
 
There is another useful (reducible) representation of the TL semigroup in   
a spin basis. The matrices
\begin{equation}\label{2.14}
e_j = \frac{1}{2}[
\sigma_j^x\sigma_{j+1}^x
+ \sigma_j^y\sigma_{j+1}^y
-\frac{1}{2} \sigma_j^z\sigma_{j+1}^z
+i\frac{\sqrt{3}}{2}(\sigma_j^z - \sigma_{j+1}^z)],
\end{equation}
where $\sigma^x$,$\sigma^y$ and $\sigma^z$ are Pauli matrices, satisfy \rf{2.10}. The 
Hamiltonian given by \rf{2.11} becomes a spin 1/2 quantum spin chain acting in 
a $2^L$ dimensional vector space. It is known that this  reducible 
representation contains all the irreducible representations of the TL 
algebra, in particular the one given by the left ideal discussed above.

\section{The raise and peel model and conformal invariance} \label{sect3}

 The quantum spin chain defined by the Hamiltonian \rf{2.11} and \rf{2.14} 
commutes not only with
\begin{equation} \label{3.1}
S^z = \frac{1}{2}\sum_{j=1}^L \sigma_j^z,
\end{equation}
but with two other operators $S^+$ and $S^-$ (see \cite{PS}). The three operators 
$S^z$, $S^+$ and $S^-$ which are related by commutation relations define the 
quantum algebra $U_q (sl(2))$ for $q = \exp(i\pi/3)$. This is a deformation of 
the usual angular momentum $sl(2)$ algebra with irreducible 
representations of dimension $2s + 1$ labeled by the spin $s$ (integer or half 
integer). The number of scalars ($s=0$ representations) for a chain with $L = 2n$ 
sites is given by the Catalan numbers \rf{2.13} which coincide with the 
number of states in the left ideal given by \rf{2.12}. It can be shown that 
this is not a simple coincidence. The spectrum of the Hamiltonian defined 
in this left ideal is obtained considering only the scalar sector of the 
quantum spin chain.

 The quantum spin chain is an integrable system \cite{AAA}. This implies that 
it's spectrum can be computed exactly using the Bethe ansatz.  If we denote by $E_r$ ($r = 
0,1\ldots, 2^L$) the energy levels in nondecreasing order: $E_0 =0 < E_1 \leq E_2 \leq \cdots$, 
the finite-size scaling partition function  of $H$ is defined as follows :
  \begin{equation} \label{3.2}
Z(q) = \lim_{L \to \infty} Z_L(q) = \lim_{L\to \infty} \sum_n q^{LE_n/\pi v_s},
\end{equation}
where $v_s = 3\sqrt{3}/2$ is the sound velocity \cite{AAA}.
 One can show \cite{BS} that $Z(q)$ 
has the expression
\begin{equation} \label{3.3}
Z(q) = \sum_s (2s+1)\zeta_s(q).
\end{equation}
Here $s$ is the spin, taking the values $s = 0,1,2,\ldots$, for $L$ even, and
$s =\frac{1}{2},\frac{3}{2},\frac{5}{2},\ldots$, for $L$ odd, and 
\begin{equation} \label{3.4}
\zeta_s(q) = q^{\Delta_s} (1- q^{2s+1}) \prod_{n=1}^{\infty}(1-q^n)^{-1},
\end{equation}
where
\begin{equation} \label{3.5}
\Delta_s = \frac{s(2s-1)}{3}.
\end{equation}

 This implies that  for large lattice sizes, the energies are (see \rf{3.3} and \rf{3.4})
\begin{equation} \label{3.6}
E = \frac{3\pi \sqrt{3}}{2L}(\Delta_s+k) + o(\frac{1}{L}),
\end{equation}
where $k$ is an integer. One can observe that the finite-size scaling 
spectrum of the Hamiltonian is expressed in terms of characters of a 
Virasoro algebra \cite{CI} with a central charge $c = 0$ (the ground-state 
has energy zero without finite-size corrections). Hence the system is conformally 
 invariant. 
 This implies a complete knowledge of the finite-size 
scaling spectrum. In the spin zero sector, for $L$ even, one gets for the first 
excitations (see Eqs. \rf{3.5}-\rf{3.6}):
\begin{equation} \label{3.7}
E_1 = 3\pi\sqrt{3}/L, \;\; E_2= \pi\frac{9\sqrt{3}}{2L},\ldots ,
\end{equation}
which correspond to the values $\Delta_s=0$ and $k=2,3,\ldots$ in Eq.~\rf{3.6}.

 In a conformal theory the 
$\Delta_s$ given by Eq. \rf{3.5} are related to the 
critical exponents of various correlation functions \cite{CI}. Notice that
\begin{eqnarray} \label{3.8}
\Delta_1 &=& \frac{1}{3}, \; \Delta_2 = 2 \;\;\;\;\;(L \;\;\mbox{even})
\nonumber \\
\Delta_{\frac{3}{2}} &=& 1, \; \Delta_{\frac{5}{2}} = \frac{10}{3}\;\;\;\;\;
(L\;\; \mbox{odd}).
\end{eqnarray}

 Eqs. \rf{3.7} and \rf{3.8} will be used in the next sections.
 The dynamical critical exponent $z$ of a stochastic process is defined by 
the finite-size scaling behaviour of the Hamiltonian spectrum:
\begin{equation} \label{3.9}
\lim_{L\to \infty} L^zE = \mbox{const}.   
\end{equation}
Comparing Eqs. \rf{3.6} and \rf{3.9} we conclude that in the RPM, $z = 1$.

\section{The raise and peel model:  combinatorial facets}\label{sect4}

 The TL semigroup algebra \rf{2.10}:
\beq
e_j^2 = e_j,\qquad e_je_{j\pm1}e_j = e_j, \qquad e_je_k = e_ke_j \quad
{\rm for} \quad |j-k| > 1,
\label{4.1}
\eeq
can be understood in terms of graphs \cite{PM}. The generators $e_j$ can be 
pictorially represented by
\beq
e_j \quad = \quad
%\begin{picture}(140,15)
\begin{picture}(140,20)
\put(2,-10){\epsfxsize=135pt\epsfbox{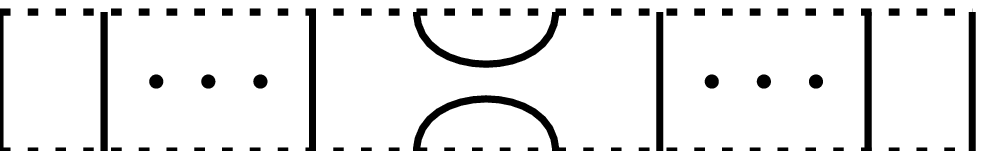}}
\put(.5,-18){$\sstyle 1$}
\put(15,-18){$\sstyle 2$}
\put(38,-18){$\sstyle j-1$}
\put(58,-18){$\sstyle j$}
\put(72,-18){$\sstyle j+1$}
\put(89,-18){$\sstyle j+2$}
\put(112,-18){$\sstyle L-1$}
\put(135.5,-18){$\sstyle L$}
\end{picture}
\hspace{10pt}\vspace{28pt}
\label{4.2}
\eeq
\vspace{1cm}

 The words in the left ideal discussed in Section 2 (Eq. \rf{2.12}) can be 
represented by boundary diagrams of loops or link patterns \cite{PM}. An example 
of such a diagram is shown in Fig. 1. If the TL semigroup has $L-1$ 
generators one takes $L$ sites. Contour lines connect pairs of sites 
and don't intersect.
%%%%%%%%%%%%%%%

\begin{figure}[ht]
\centerline{
\begin{picture}(270,40)
\put(50,0){\epsfxsize=160pt\epsfbox{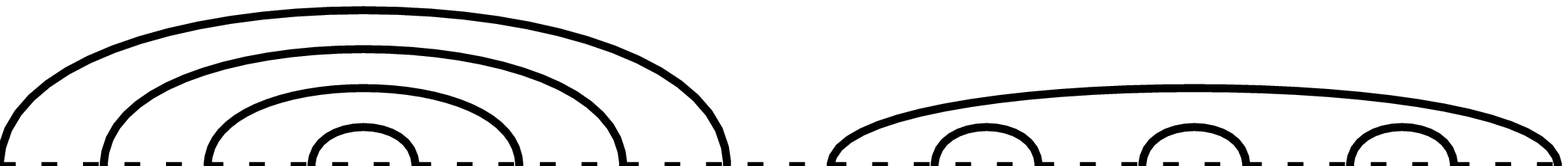}}
%\put(0,0){\epsfxsize=160pt\epsfbox{fig1loops.eps}}
%\put(180,0){$\leftrightarrow$}
%\put(210,0){\epsfxsize=160pt\epsfbox{conn.eps}}
\end{picture}}
\caption{An example of a link pattern for $L=16$}
\label{Ifig1}
\end{figure}
%%%%%%%%%%%%%%%%%%  

 The action of $e_j$ on a link pattern of contour lines is given by placing 
the graph of $e_j$ underneath that of the link pattern one, removing the 
closed loops and the intermediate dashed line. Next, one contracts the 
links in composite pictures. The action of $e_1$ on one of the link patterns 
for $L = 6$ is given by,
\beq \label{4.3}
\begin{picture}(270,40)
\put(0,0){\epsfxsize=120pt\epsfbox{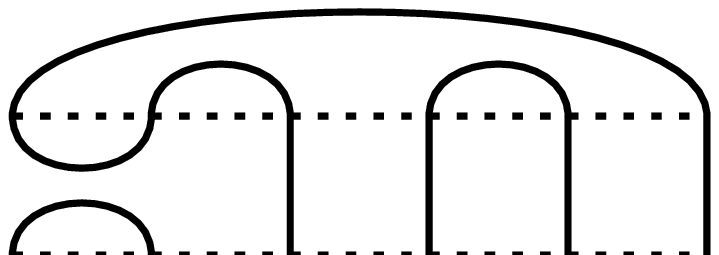}}
\put(130,10){$=$}
\put(150,10){\epsfxsize=120pt\epsfbox{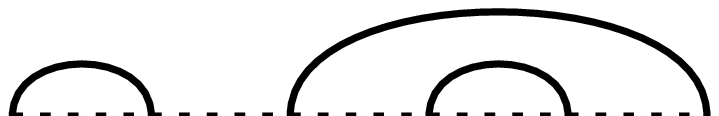}}
\end{picture}
\eeq

 We consider the action of the Hamiltonian
\beq     
                   H = \sum_{j=1}^{L-1} (1-e_j),
\label{4.4}                                                
\eeq
on the link patterns. As an example we consider the case $L = 6$. In this 
case we have five words in the ideal $I_1$: 1) $J_0 = e_1e_3e_5$, 2) $e_2J_0$, 
3) $e_4J_0$, 4) $e_2e_4J_0$ and 5) $e_3e_2e_4J_0$. These words correspond to 
the link patterns:

\beq
\begin{picture}(240,80)
\put(0,0){\epsfxsize=240pt\epsfbox{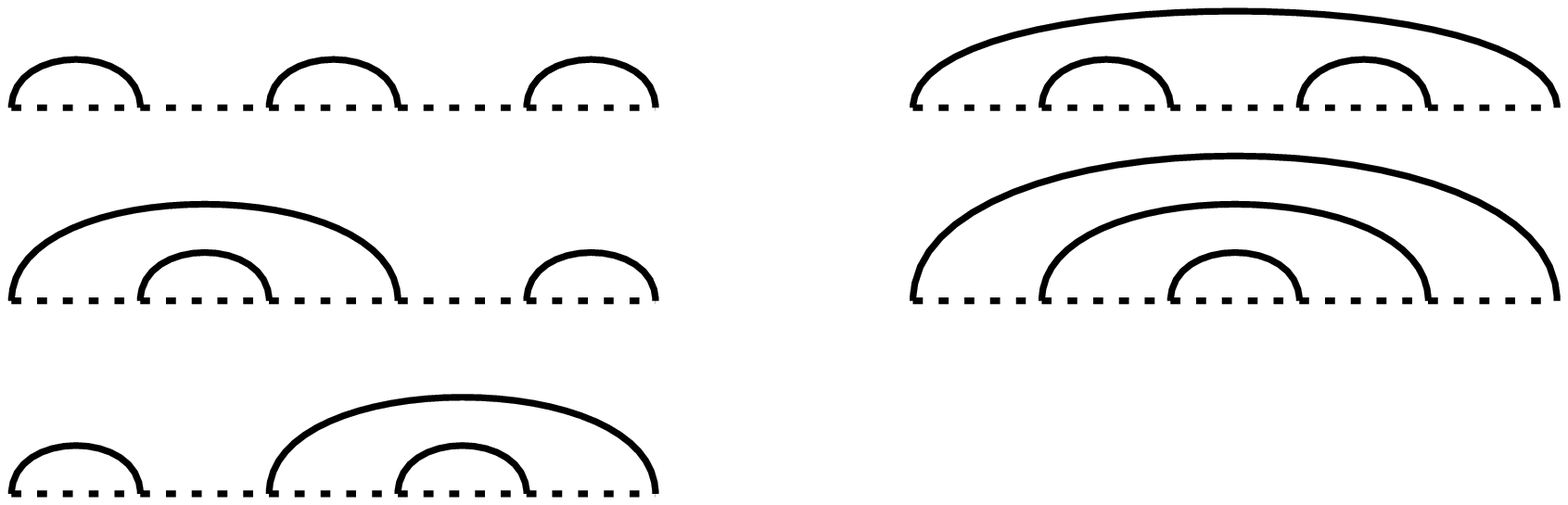}}
\put(-20,60){$1:$}
\put(-20,30){$2:$}
\put(-20,0){$3:$}
\put(120,60){$4:$}
\put(120,30){$5:$}
\end{picture}
\label{4.5}
\eeq
and we find,
\beq \label{4.6}
H = - \left( \begin{array}{@{}ccccc@{}}
-2 & 2 & 2 & 0 & 2 \\
1 & -3 & 0 & 1 & 0 \\
1 & 0 & -3 & 1 & 0 \\
0 & 1 & 1 & -3 & 2 \\
0 & 0 & 0 & 1 & -4
\end{array}\right).
\eeq
Which is indeed an intensity matrix. In the basis \rf{4.5} the stationary 
state $\ket0$ of $H$ is given by
\beq
\ket0 = (11,5,5,4,1).
\label{4.7}
\eeq
Notice that all the components of $\ket0$ are integer numbers (we have chosen 
the smallest component to equal 1). The normalization factor $\bra0 \ket0$ (equal 
to the sum of the components of $\ket0$) is equal to 26. Below we show that 
the normalization factor acquires an extra meaning from an enumeration 
problem \cite{BGN}: it is equal to the partition function of an equilibrium 
statistical mechanics system in two dimensions. This observation has deeper consequences 
which we will now briefly discuss.

\begin{figure}[ht]
\centerline{
\begin{picture}(240,80)
\put(0,0){\epsfxsize=80pt\epsfbox{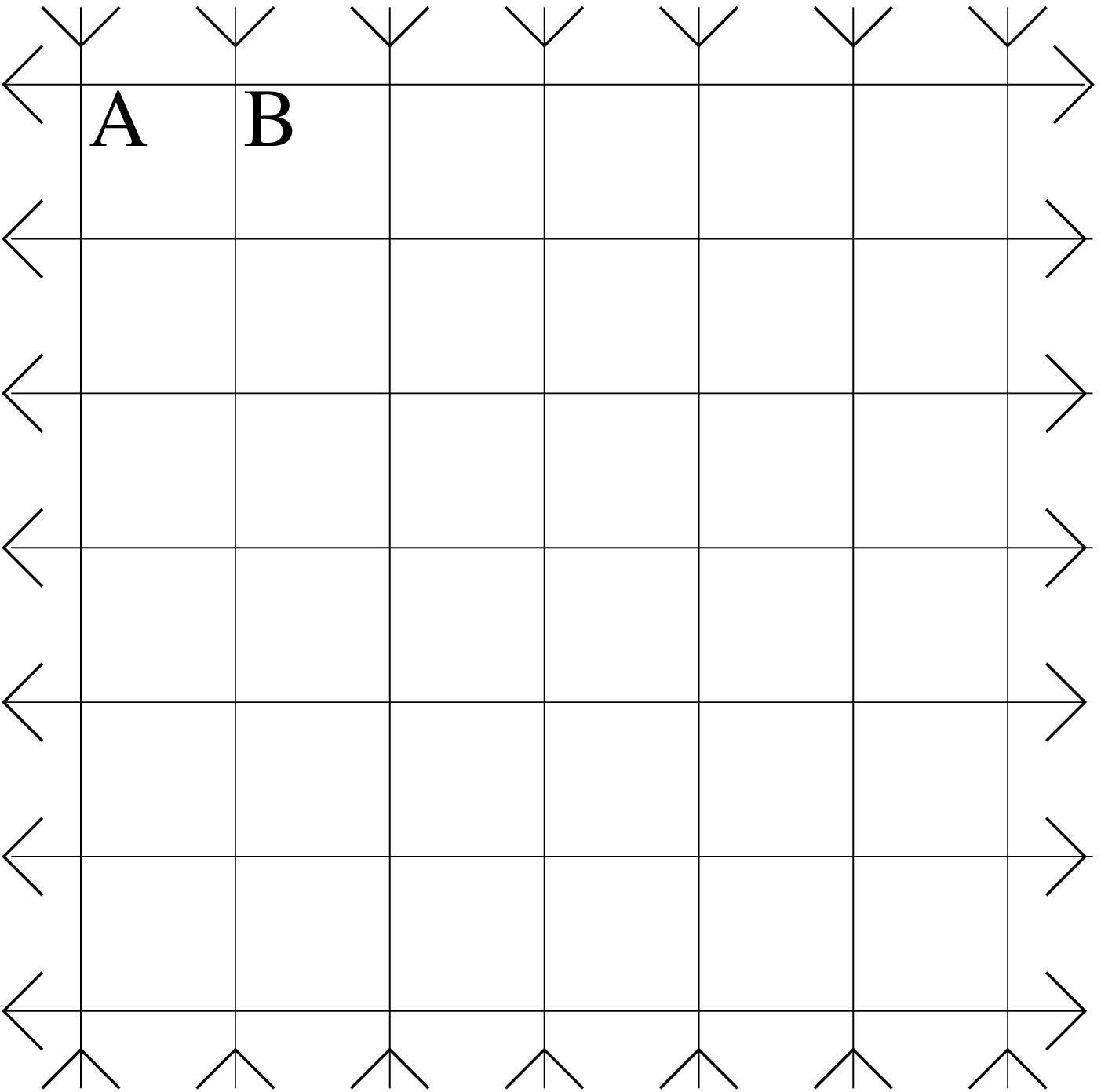}}
\put(90,40){(a)}
%\put(120,40){$\rightarrow$}
\put(160,0){\epsfxsize=80pt\epsfbox{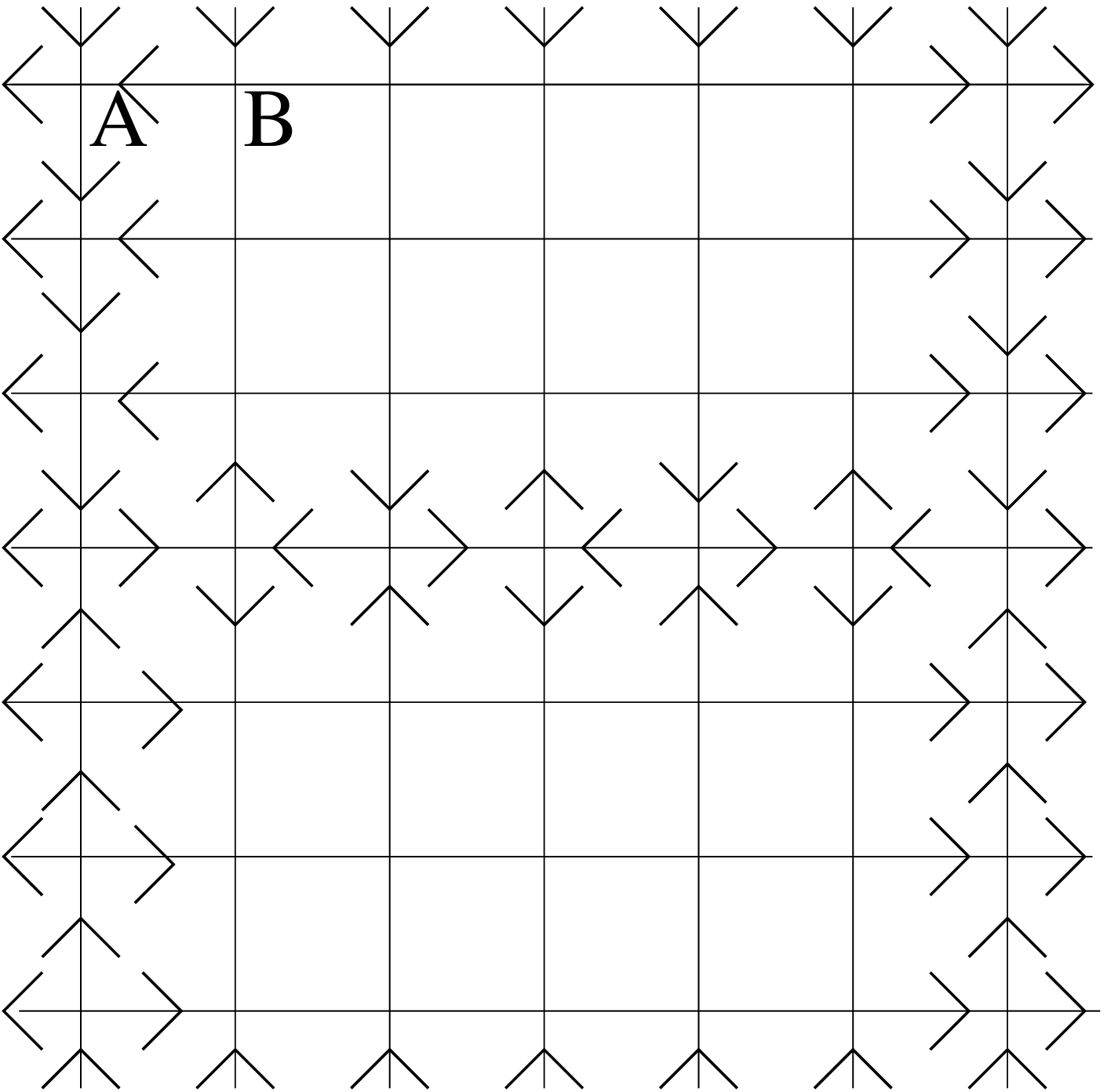}}
\put(260,40){(b)}
\end{picture}}
\caption{
a) Domain wall boundary condition for the six-vertex model. 
b) Fixed arrows for the horizontally symmetric vertex configurations. The lattice size is $(L+1)\times (L+1)$ with $L=6$.}
\label{fign1}
\end{figure}

\begin{figure}[ht]
\centerline{
\begin{picture}(330,170)
\put(30,0){\epsfxsize=300pt\epsfbox{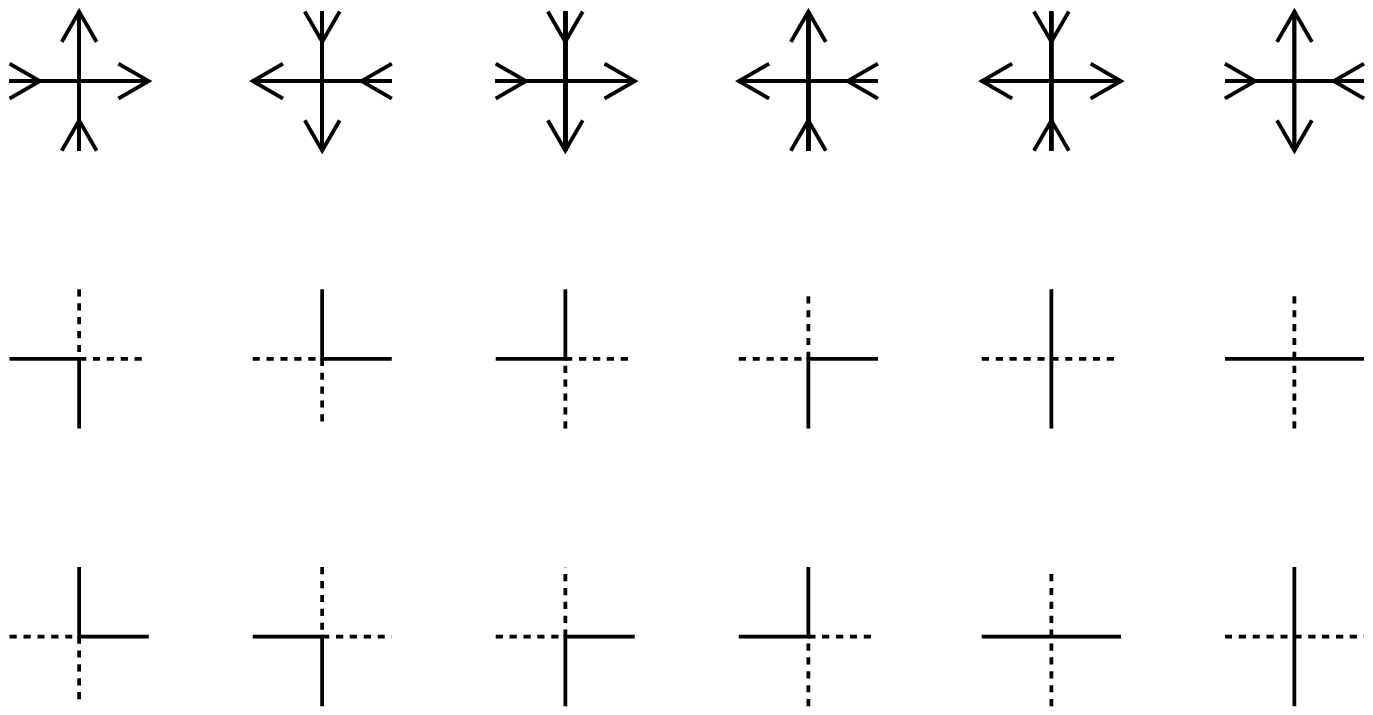}}
\put(0,78){A}
\put(0,20){B}
\end{picture}}
\caption{FPL vertices on sublattices A and B derived from the six arrow vertices.}
\label{fig3}
\end{figure}

We consider a six-vertex model \cite{SIX} in a $(L+1)\times (L+1)$ square lattice with 
{\it domain wall boundary condition} \cite{KO}. This boundary condition 
imposes that all the arrows on the vertical (horizontal) boundaries point outward 
(inward). In Fig.~\ref{fign1}(a) we show the fixed links in this  boundary condition 
for the case $L=6$. 
The arrow configurations on this lattice can be transformed 
	 into FPL configurations. The latter are configurations of 
paths such that every site is visited by exactly one path. 
This transformation is done by dividing  the 
square lattice into its even and odd sublattices denoted by A and B,  
respectively. Instead of arrows, only those edges are drawn that on 
sublattice A point inward and on sublattice B point outward as in 
Fig.~\ref{fig3}.
 We take the vertex in the upper left corner to belong to sublattice A. 
%%%%%%%%%%%%%%
 
\begin{figure}[ht]
\centerline{
\begin{picture}(240,80)
\put(0,0){\epsfxsize=80pt\epsfbox{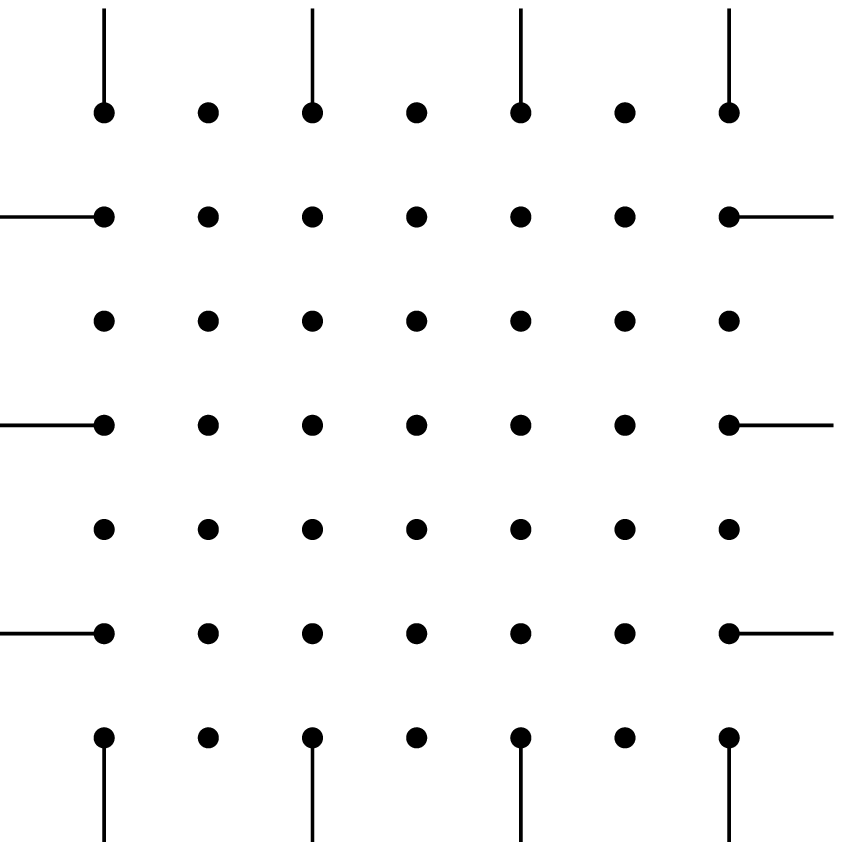}}
\put(90,40){(a)}
\put(160,0){\epsfxsize=80pt\epsfbox{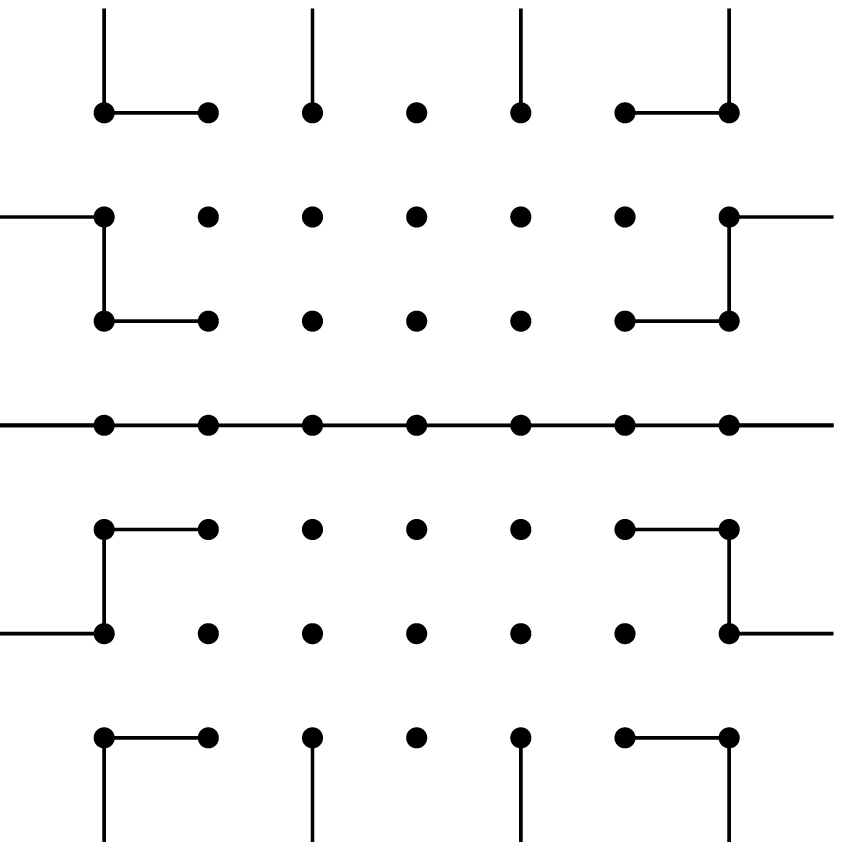}}
\put(260,40){(b)}
\end{picture}}
\caption{
a) Boundary conditions (fixed edges) for the FPL diagrams .
b) Fixed edges for the horizontally symmetric FPL diagrams. The lattice size is $(L+1)\times (L+1)$ with $L=6$.}
\label{fig4}
\end{figure}

%%%%%%%%%%%%%%%%%%%%%%%%%%%
The domain wall boundary conditions of the six-vertex model translates 
into a boundary condition for loops. For example, the boundary condition 
shown in Fig.~\ref{fign1}(a) translates into the boundary conditions given in 
Fig.~\ref{fig4}(a). 
%%%%%%%%%%%%%%%%%%%%%%%%%%%%%%%%%%%%%%%%%%%%%%%%%%%

%%%%%%%%%%%%%%%%%%%%%%%%%%%%%%%%%%%%%%%%%%%%%%%%%%%%%%%%%%%

Among the general configurations of the vertex model, we now restrict 
ourselves to those which are horizontally symmetric. 
In the case of $L=6$ the boundary condition is shown in 
Fig.~\ref{fign1}(b). Due to the correspondence of vertices given in 
Fig.~\ref{fig3} the related FPL configurations  will also be horizontally 
symmetric. In 
Fig.~\ref{fig4}(b) we show the boundary condition for FPL related with that 
of Fig.~\ref{fign1}(b), for $L=6$. 

\begin{figure}[ht]
\centerline{
\begin{picture}(90,50)
\put(0,0){\epsfxsize=90pt\epsfbox{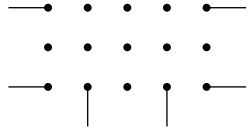}}
\end{picture}}
\caption{ FPL grid corresponding to Fig. 2.}
\label{fig5}
\end{figure}

The above relations show that number of horizontally symmetric configurations 
of the six-vertex model in the $(L+1)\times (L+1)$ square lattice with 
domain wall boundary condition is the same  as 
the number of FPL diagrams in the $(L-1)\times L/2$ rectangular lattice. The 
boundary condition for the FPL diagrams is special, for the case $L=6$ is 
shown in Fig.~\ref{fig5}. 
 
%%%%%%%%%%%%%%%%%%%%%%%%%%%%%%%%%%%%%%%%%%

 The paths that start and end on boundary sites (we disregard the closed 
loops in the bulk of the FPL diagrams, as a consequence of the 
correspondence given in Fig.~\ref{fig3}) define link patterns having the 
same topology as the TL link patterns \rf{4.5}. 
There is a fascinating 
connection between the components of the stationary state $\ket0$ 
(see 
Eq. \rf{4.7}) and the enumeration of FPL configurations on the rectangular 
grid: the unnormalized probability of a state corresponding to a given link 
pattern is equal to the number of FPL configurations with the same link 
pattern \cite{RS1,PRGN}. This is an yet unproven conjecture.
 
Take for example the stationary state \rf{4.7} for $L = 6$. The FPL 
configurations on the $5\times 3$ rectangle are shown in Fig. 6. Their total 
number is 26, which is equal to $\bra0 \ket0$ and they can be categorized 
according to the five link patterns present in \rf{4.6}. One finds that 
\begin{figure}[ht]
\centerline{
\begin{picture}(340,200)
\put(0,0){\epsfxsize=340pt\epsfbox{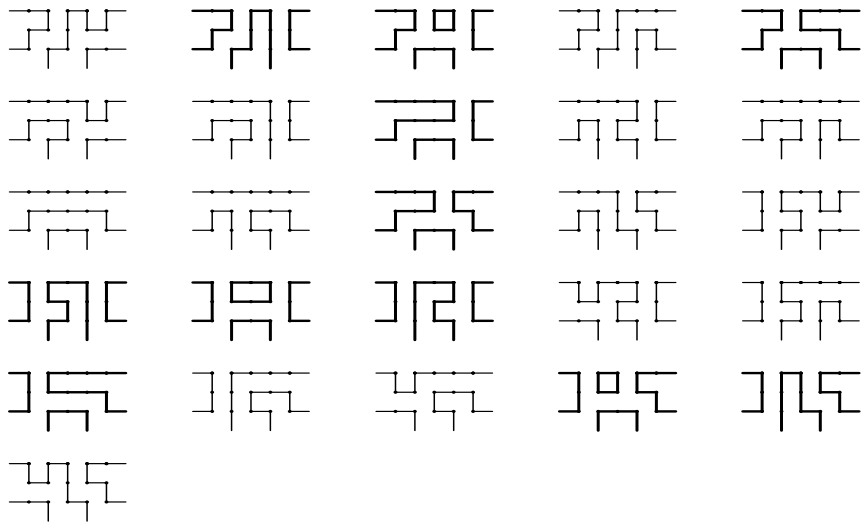}}
\end{picture}}
\caption{The $26$ FPL diagrams for $L=6$. The $11$ diagrams
corresponding to link pattern $1$ in \rf{4.5} are printed in 
bold.}
\label{fig6}
\end{figure}
the number of diagrams corresponding to the link pattern 
$1$ in \rf{4.5} is $11$ 
(they are printed in bold in Fig. 6), to link pattern $2$ is $5$, to link 
pattern $3$ is $5$, to link pattern $4$ is $4$ and to link pattern $5$ is $1$.

 We have illustrated for $L = 6$ the connection between  
 the unnormalized probabilities in the stationary state of various 
link patterns in the RPM at the Razumov-Stroganov point and the number of FPL 
configurations with the same link pattern. This connection is important 
since it shows that one can see a "far from equilibrium" stationary state 
of an one-dimensional system as an equilibrium state of a system in two dimensions. 

 Before we further elaborate on this point, we will shortly discuss 
another facet of the RPM: its connection with the enumeration 
of alternating sign matrices (ASM) (see \cite{BR} for an excellent introduction 
to this subject). 

 An alternating sign matrix is a square matrix of $0$s, $1$s and $-1$s for which
a) the sum of the entries in each row and each column is $1$.
b) the non-zero entries of each row and of each column alternate in sign.
 An example of such a matrix is
\beq \label{4.8}
\left( \begin{array}{@{}ccccc@{}}
0 & 1 & 0 & 0 & 0 \\ 
0 & 0 & 1 & 0 & 0 \\ 
1 &-1 & 0 & 0 & 1 \\          
0 & 1& -1 & 1 & 0 \\
0 & 0 & 1 & 0 & 0
\end{array}  \right).
\eeq
 
 Alternating sign matrices can have symmetry properties. For example, the 
matrix
\beq \label{4.9}
\left( \begin{array}{@{}ccc@{}}
0&  1 & 0 \\
1& -1 & 1 \\                
0 & 1 & 0
\end{array}  \right)
\eeq
is vertically and horizontally symmetric. There is a one to one 
correspondence between the six-vertex configurations with domain-wall 
boundary conditions and ASM. To each vertex you attach the numbers $0$, $-1$ 
or $1$ as shown in Fig. 7. To six-vertex configurations 
with a certain 
symmetry correspond alternating sign matrices with the same symmetry.
%%%%%%%%%%%%%%%%ARRUMAR%%%%%%%%%%%%%%%%%%%%%%%%%
\begin{figure}[ht]
\centerline{
\begin{picture}(330,70)
\put(0,0){\epsfxsize=300pt\epsfbox{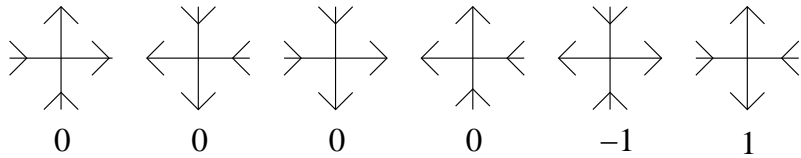}}
%\put(30,0){\epsfxsize=300pt\epsfbox{6VtoFPL.eps}}
%\put(0,78){A}
%\put(0,20){B}
\end{picture}}
\caption{Correspondence between vertices and entries of 
alternating sign 
matrices.}
\label{fig7}
\end{figure}
%%%%%%%%%%%%%%%%%%%%%%%%%%%%%%%%%%%%%%%%%%%%%%%%

 The number of six-vertex configurations on a $(L+1)\times(L+1)$
 square with 
domain wall boundary conditions that are invariant under reflection in the 
horizontal symmetry axis is equal to that of the $(L-1)\times L/2$ rectangle with 
special boundary conditions like in Fig.~\ref{fign1}(b). The total number of such 
configurations is known and is equal to the number of $(L+1)\times (L+1)$ 
horizontally symmetric ASM (equal to the number of vertically symmetric 
ASM) which is given by \cite{KU}: 
\beq
A^{\rm V}_{2n+1} = \prod_{j=0}^{n-1} (3j+2)
\frac{(2j+1)!(6j+3)!}{(4j+2)!(4j+3)!} = 1,3,26,646,\ldots, 
\label{4.10}
\eeq
where $L = 2n$. The leading asymptotic terms of $A_{L+1}^V$
 are given by
\beq
\ln A^{\rm V}_{L+1} = s_0 \frac{(L-1)L}{2} + 
(L-1)\ln \frac{3\sqrt{6}}{8} - \frac{5}{144}
%\ln A^{\rm V}_{L+1} = s_0 \frac{(L-1)L}{2} + (s_0-\frac12\ln 2) (L-1) - \frac{5}{144}
%\ln L^2 + O(1),\;\; s_0 = \ln (\frac{3\sqrt{3}}{4}). \nonumber 
\ln L^2 + O(1),
\label{4.11}
\eeq
where  $s_0 = \ln (\frac{3\sqrt{3}}{4})$.
The first term is proportional to the area of the rectangle, the second 
one is related to the boundary. This shows that, indeed, the stationary 
state of the RPM is related to a {\it bona fide} two-dimensional statistical
mechanics system. 

 In the next section we are going to discuss physical applications of the
stochastic model defined by the Hamiltonian $H$ (Eq.~\rf{4.4}) 
acting on the
vector space of link patterns (see, for $L=6$, Eq.~\rf{4.5}).
 In the Appendix  we show how
each link pattern can be mapped onto a RSOS (Dick) path. 
Such a path can 
be
seen as an one-dimensional interface separating clusters of "tiles"
deposited on a substrate, from a rarefied gas of tiles. The action of the
Hamiltonian $H$ on the RSOS configurations gives the transition rates for
various adsorption and desorption processes (see Eqs.~\rf{4.5}, \rf{4.6} and 
Fig.~\ref{fignova} for an example with $L=6$).

\section{The raise and peel model: the phase diagram}
\label{sect6}

We 
 consider a one-dimensional lattice with $L+1$  
($L=2n$) sites.  An  interface is   formed by attaching at each site 
non-negative integer heights $h_i$ ($i=0,1,\ldots,L$) which obey the restricted
solid-on-solid (RSOS) rules:
\beq
h_{i+1} - h_i = \pm1,\qquad h_0 = h_L = 0,\qquad h_i \geq 0.
\label{I1}
\eeq

There are $C_n= (2n)!/((n+1)(n!)^2)$ possible configurations of the
interface, that also correspond to the number of independent words \rf{2.13} in the left ideal 
$I_1$  given in \rf{2.12} (see Sec. 4). In Fig.~\ref{Ifig1} we show a configuration for
$n=8$ ($L=16$).
\begin{figure}[h]
\centerline{
\begin{picture}(160,40)
\put(0,0){\epsfxsize=160pt\epsfbox{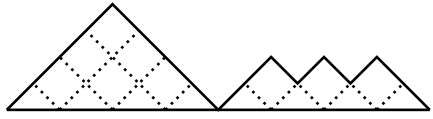}}
\end{picture}}
\caption{A configuration of the interface with three contact points and
two clusters, as defined in Sec.~\ref{sect6a}, for the lattice size $L=16$.}
\label{afig1}
\end{figure}
Alternatively, one can describe the interface using slope variables
$s_i = (h_{i+1} - h_{i-1})/2$,  ($i=1,...,L-1$).

The dynamics of the interface is described in a transparent way in the
language of tiles (tilted squares) which cover the area between the
interface and the substrate $(h_{2i}=0$, $h_{2i+1} =1$,
$(i=0,...,n-1), h_L=0)$ (see Fig.~\ref{afig1}).

\begin{figure}[t]
\centerline{
\begin{picture}(160,80)
\put(0,10){\epsfxsize=160pt\epsfbox{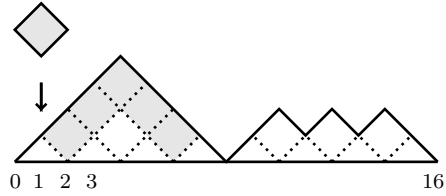}}
\put(-2,0){$\sstyle 0$}
\put(7,0){$\sstyle 1$}
\put(17,0){$\sstyle 2$}
\put(27,0){$\sstyle 3$}
\put(154,0){$\sstyle 16$}
\end{picture}}
\caption{A desorption event. The incoming tile at site $1$ triggers, 
with probability $u_d$,  an
avalanche of $5$ tiles, which are shaded. All of the shaded tiles are
removed in the desorption event.}
\label{afig2}
\end{figure}

We consider the interface separating a film of tiles
deposited on the substrate from a rarefied gas of tiles. 
The evolution of the system in discrete time (Monte-Carlo steps) is
given by the following rules. With a probability $P_i= 1/(L-1)$ a tile from
the gas hits site $i,\; (i=1,...,L-1)$. Depending on the value of the slope
$s_i$ at the site $i$, the following processes can occur:
\begin{itemize}
\item[i)] $s_i=0$ and $h_i > h_{i-1}$. %(local maximum)

The tile hits a local peak and is reflected.
\item[ii)] $s_i=0$ and $h_i < h_{i-1}$. %(local minimum)

The tile hits a local minimum. With a probability $u_{\rm a}$ the tile
is adsorbed ($h_i \mapsto h_i+2$) and with a probability $1-u_{\rm
a}$ the tile is reflected.
\item[iii)] $s_i=1$.

With probability $u_{\rm d}$ the tile is reflected after triggering
the desorption of a layer of tiles from the segment
($h_j>h_i=h_{i+b},\; j=i+1,\ldots, i+b-1$),
{\it i.e.} $h_j \mapsto h_j-2$ for $j=i+1,...,i+b-1$. This layer contains
$b-1$ tiles (this is always an odd number). With a probability $1-u_{\rm d}$, the
tile is
reflected and no desorption takes place. For an example see Fig.~\ref{afig2}.
\item[iv)] $s_i=-1$.

With probability $u_{\rm d}$ the tile is reflected after triggering
the desorption of a layer of tiles belonging to the segment
($h_j>h_i=h_{i-b},\; j=i-b+1,\ldots, i-1$),
{\it i.e.}  $h_j \mapsto h_j-2$ for $j=i-b+1,...,i-1$. With
a probability $1-u_{\rm d}$ the tile is reflected and no desorption
takes place.
\end{itemize}

The physics of the model depends on one parameter $u=1/w=u_a/u_d$. 
%insertion in the revise version
% parei
We use these notations since we consider
below either very small values of $w$ (or $u$) and our intuition can
be helped by the fact that for $u=0$ or $w=0$ the properties of the
system are known. Namely, for both $u=0$ and $w=0$ the spectrum is
massive \cite{GNPR1}, the stationary states being the substrate ($u=0$) and a
full triangle ($w=0$), i. e.,  $h_i=h_{L-i}$ ($i=1,\ldots,L/2$).
%end of insertion for revision  version

At  the special point of the RPM where the ratio $u = u_a/u_d=1/w=1$ (the 
Razumov-Stroganov point considered in earlier sections) the model is
governed by the Temperley-Lieb Hamiltonian \rf{4.4} acting on the 
space of configurations  by the ideal $I_1$ defined in 
\rf{2.12} (see Sec. 4 and 
Appendix). As a consequence, the model is exactly 
integrable and conformally invariant. 
As discussed in Sec. 3 this last invariance 
allows exact predictions for the finite-size eigenspectrum of $H$.
For $u\neq 1$ the model is not exact integrable anymore and our knowledge comes from 
from Monte Carlo simulations \cite{ALR}. 
%For $u=0$ and $w=0$ the spectrum is 
%massive \cite{GNPR1}, the stationary states being the substrate ($u=0$) and full triangle 
%($w=0$).

\begin{figure}[t]
\centerline{
\begin{picture}(300,50)
\put(0,15){\epsfxsize=300pt\epsfbox{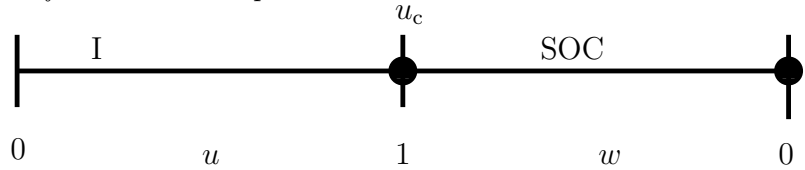}}
\put(-0.5,2){$0$}
\put(290,0){$0$}
\put(145,0){$1$}
\put(72,0){$u$}
\put(222,0){$w$}
\put(145,55){$u_{\rm c}$}
\put(30,40){I}
\put(200,40){SOC}
\end{picture}}
\caption{
     Phase diagram of the raise and peel model. For $0 \leq u < 1$ (phase I),
     the
    model is massive. At $u = u_c=1$ it is massless and conformal invariant.
    For
    $1 > w >0 $ it is scale invariant with varying critical exponents and
    exhibits  SOC.
     For $w = 0$ the system is massive. }

\label{afig4}
\end{figure}

The 
    phase diagram of the model obtained from the Monte Carlo simulations is given in Fig.~\ref{afig4}.
     In the domain $0 < u< 1$ the system is massive undergoing a
    second order phase transition at $u = u_c = 1$. For $1 > w > 0$
    the conformal symmetry seen at
    $w= 1$ is broken but the system stays scale invariant. Since in this
    domain one has
    avalanches we call this phase a self-organized criticality (SOC) phase. Another phase transition
    occurs at $w = 0$
    where the system becomes massive again.
The phase diagram of Fig.~\ref{afig4} was established from the static and dynamic properties of the RPM \cite{ALR}. In the next section we are going to consider the 
static properties of the stationary state and in Section 7 the dynamical ones. 

\section{The stationary state of the RPM} \label{sect6a}

 It is useful to define some quantities which characterize the surface.
An obvious geometric observable
 is the set of sites $j$ ($j$ even) for which $h_j=0$
(the sites $0$ and $L$ always belong to this set). These sites are
also called contact points. This set is important, for example,  to
study  desorption. Desorption events are limited to the area between
two contact points, defined as a cluster (see Fig.~\ref{afig1}). 
 The density of contact points gives
a local observable for which one can define various correlation
functions.

 We first define quantities in the stationary state. The average number of
 clusters $k(L)$,   the density of clusters $\rho(L)$ and the local 
density of clusters $g(j,L)$ are defined by:
\beq
k(L) = \sum_{j=1}^L g(j,L)= \langle \sum_{j=1}^{L} \delta_{h_{j},0}\rangle ,
\label{I2}
\eeq
\beq
\rho(L) = k(L)/L.
\label{I3}
\eeq
  Note that the maximum value of $\rho(L)$ is $\frac{1}{2}$
  which is obtained when $u=0$
  and one has only one configuration in the system:  the substrate.
   The average height is defined as:
\beq
h(L) = \frac{\langle \sum_{j=1}^{L} h_j\rangle}{L}.
\label{I4}
\eeq
The lowest value of $h(L)$, corresponding to the substrate, is $\frac{1}{2}$.
    It is useful to consider also the average height in the middle of the
    system:
\beq
h_{\frac{1}{2}}(L) = \langle h_{\frac{L}{2}}\rangle .
\label{I5}
\eeq
     A relevant quantity is the average of the fraction of the interface where
     desorption does not take place (FND)
\beq
n(L) = \frac{1}{L-1}\langle \sum_{j=1}^{L-1}(1-|s_j|)\rangle .
\label{I6}
\eeq
      This quantity  allows to estimate the
      average number of  tiles desorbed in avalanches (see Sec.~7). 
By studying the behaviour of $n(L)$ for large $L$ and various values of $w$ we 
will be able to establish the existence of a discontinuous phase transition 
at $w=0$.

       The large $L$ behavior of any of the quantities \rf{I2}-\rf{I6}, 
say $h(L)$, will be
       denoted $h_{\infty}$:
\beq
 h_{\infty}\equiv  \lim_{L\rightarrow \infty}  h(L).
 \label{I7}
 \eeq

        In the next section we are going to discuss average quantities which are
    not only $L$ dependent but also time dependent. For these
    quantities we use, for example,  the notation $k(t,L)$ to denote the 
number of clusters.

\begin{figure}[t]
\centerline{
\begin{picture}(260,190)
\put(0,0){\epsfxsize=260pt\epsfbox{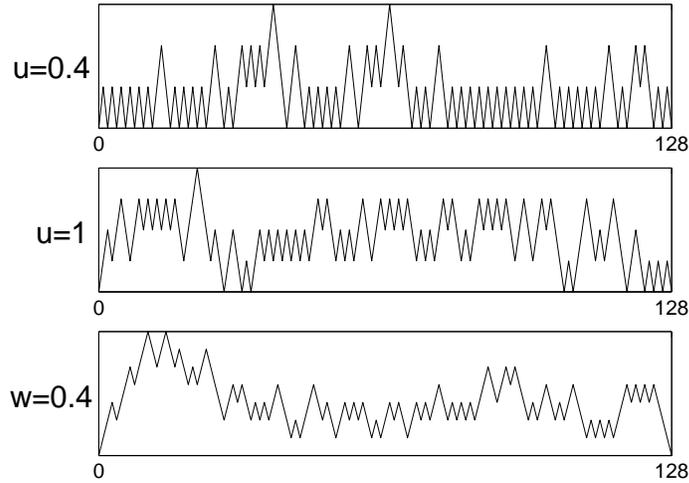}}
\end{picture}}
\caption{
  Typical configurations in the stationary states for
$ L= 128$ sites and three  values of~$u$.}
\label{afig3}
\end{figure}

 In order to get an intuition on the behavior of the model when $u$ (or $w=1/u$) 
 changes, in Fig.~\ref{afig3} we show, for $L =  128$, typical configurations in
 the
 stationary states for three values of the parameter $u$.
   One notices that for $u = 0.4$ there are many clusters,
  fewer for $u = 1$ and
  a single one for $w = 0.4$
(other, less probable, configurations do exhibit several clusters). We also notice that the average heights are quite small.

\begin{figure}[t]
\centering{\includegraphics [angle=0,scale=0.39] {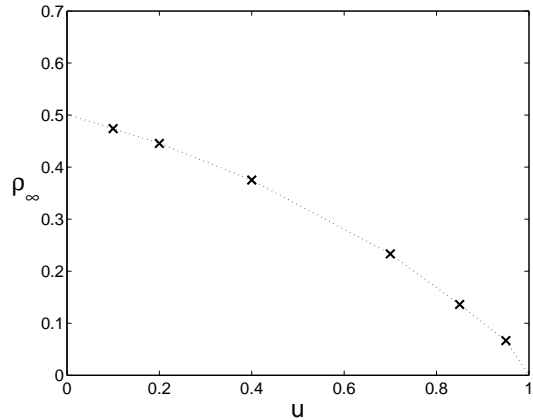}}
\caption{
        The density of clusters $\rho_{\infty}$ for various values
       of $u <  1$.}
\label{afig5}
 \end{figure}

     We first consider the density of clusters $\rho (L)$ in the domain
     $0\leq  u < 1$. Taking large values of $L$, one obtains the $u$
     dependence of
     $\rho_{\infty}$ shown in Fig.~\ref{afig5}. In all this domain the density
     of clusters
     stays finite. The density of clusters decreases from its maximum
     possible value $1/2$ which corresponds to the substrate at $u = 0$
     to the value
     zero for $u = 1$. This indicates that there is a phase transition
     at $u = 1$. From the Monte Carlo simulations it was hard to obtain the
     critical exponent which gives $\rho(u)$ when $u$ approaches the value
     one.

\begin{figure}[t]
\centering{ \includegraphics [angle=0,scale=0.46]
{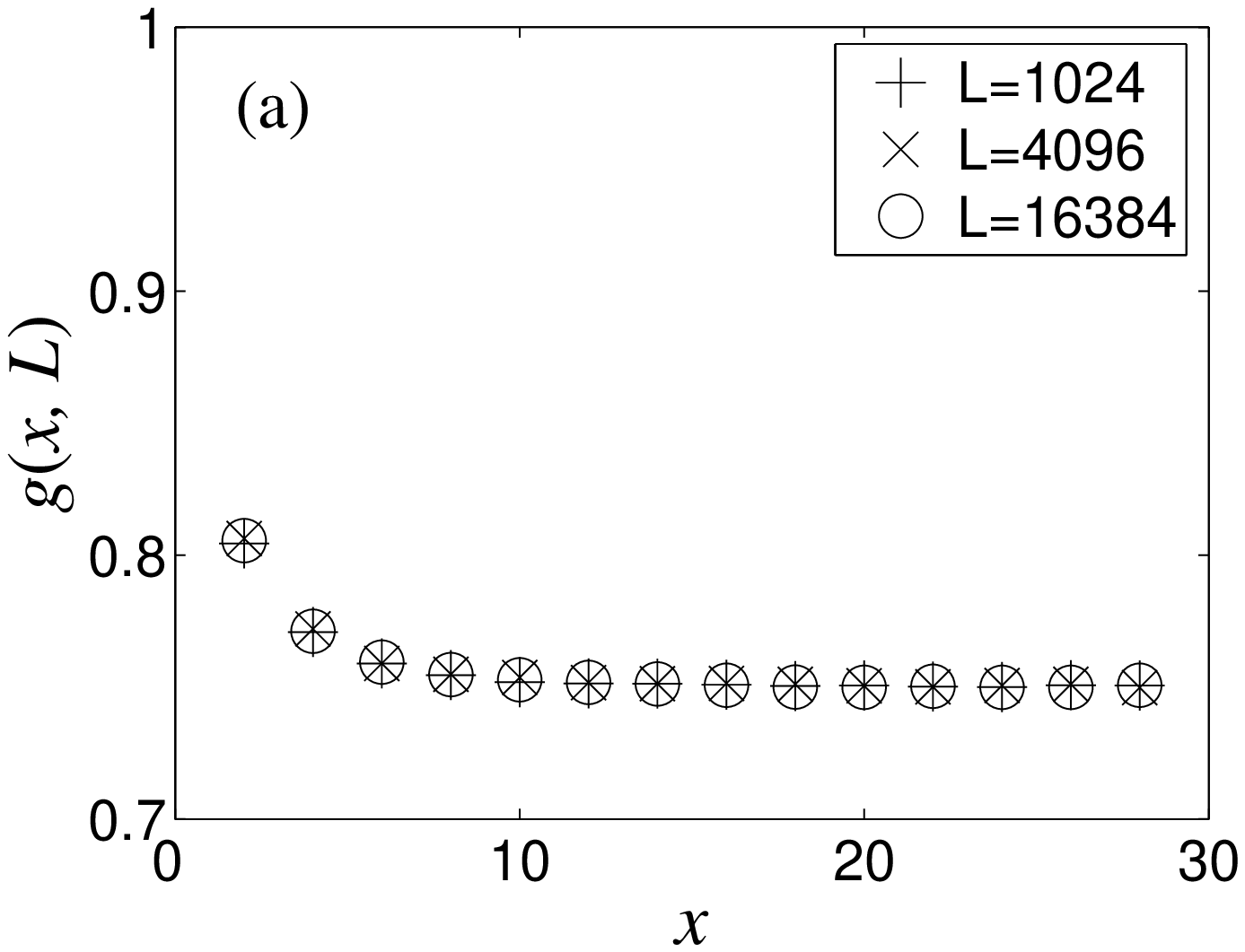}\hspace{-3.0cm}\hfill\includegraphics
[angle=0,scale=0.35] {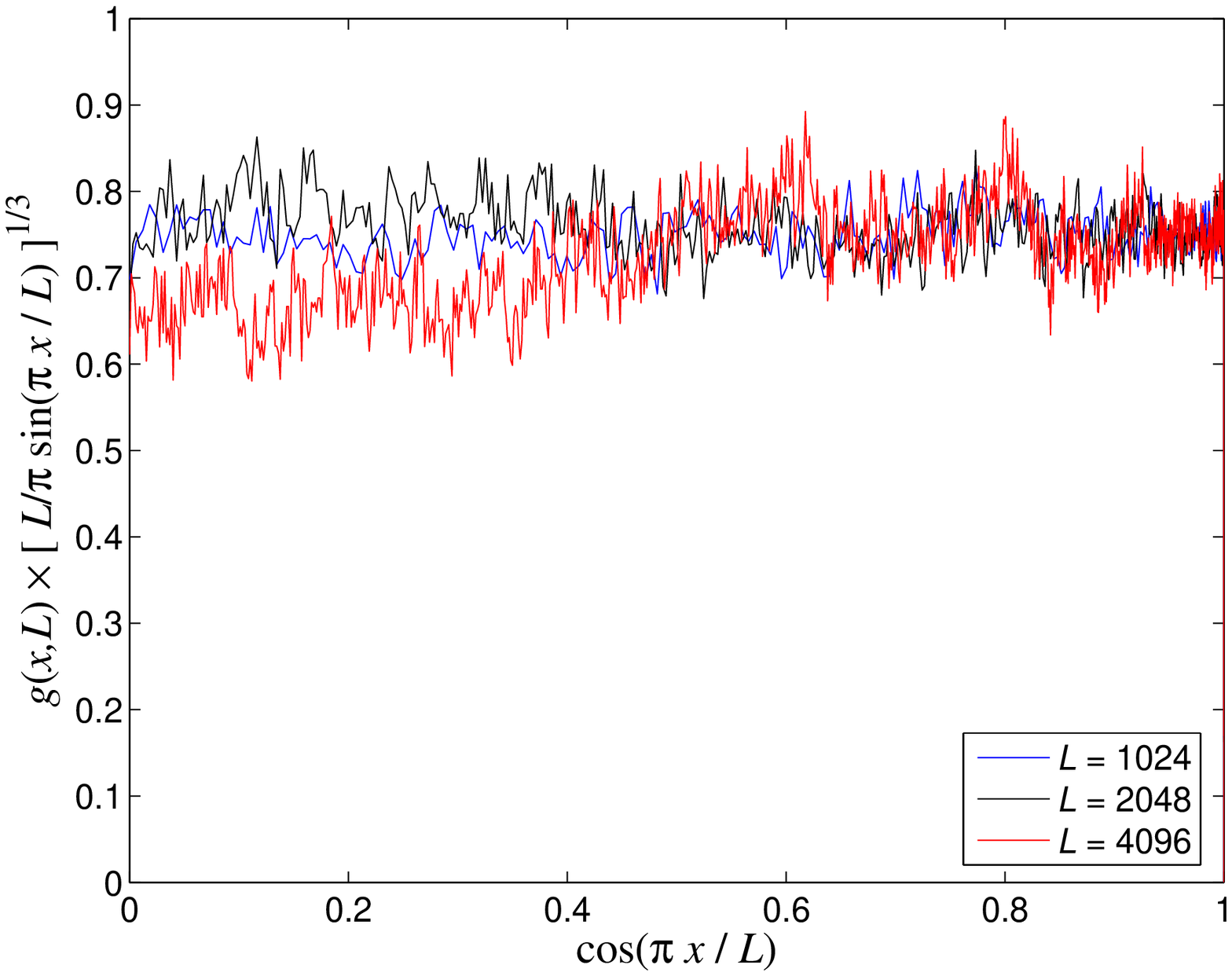} } \caption{
     (a) The local density of contact points  $g(x,L)$ at $u= 0.4$ for
    different lattice sizes $L$ ($g(x,L)$ is zero for $x$ even). (b) (Color online) The scaling function 
$C(\cos (\pi x/L))$ 
defined in (\ref{I10}) for $u=1$.}
\label{Ifig6}\label{Ifig7}
 \end{figure}

        The local density of contact points $g(x,L)$ (see Eq.~\rf{I2})
for $u = 0.4$ and different lattice sizes is shown
    in Fig.~\ref{Ifig6}(a). One
    observes that for small values of $x$, and large values of $L$,
    $g(x,L)$ decreases
    exponentially from the value 1 at $x = 0$ (not shown on Fig.~\ref{Ifig6}(a))  to a constant value
    in the bulk. 
This value is twice that seen in Fig.~\ref{afig5}, 
since for even $x$ one always have $g(x)=0$.

    Since  for $u = 0$ one can diagonalize the
     related Hamiltonian of the stochastic process and
     show rigorously \cite{GNPR1} that the system is massive, we can
     expect that
     the system is massive in the whole domain $0\leq u < 1$.

      We turn now to the interesting case $u = 1$ 
where the system is exact integrable and conformally invariant.
 In Ref.~\cite{DG} a
      conjecture was
      made which gives the probability $P_n(k)$  to have $k$ clusters in a system
      of size $L=2n$:
\beq
P_n(k) = k \frac{4^{n+k}}{27^n} \frac{\Gamma(\frac{1}{2}+n+k)\Gamma(\frac{1}{3})
\Gamma(3n+2)\Gamma(2n-k)}
{\Gamma(\frac{1}{2})\Gamma(\frac{1}{3}+2n+k) \Gamma(n+1)\Gamma(2n+k+2)\Gamma(n-k+1)} \nonumber .
\label{conj}
\eeq
The  last result gives us the  leading behaviour 
\beq 
\label{I8} k(L)= \sum_{k=1}^n kP_n(k)  \simeq \frac{\Gamma(1/3) \sqrt{3}} {2\pi}
L^{\frac{2}{3}}, 
\eeq
for the average number of clusters. This 
       shows that the density of clusters vanishes algebraically for
      $u = 1$.
This conjecture was checked using Monte Carlo simulations up
      to a system
      of size $L=512$.    Since, as we show below, the density of clusters also vanishes
      algebraically  in the entire domain $w<1$, it is convenient to define a critical exponent
      $\alpha < 1 $
 which gives the power law increase of the number
      of clusters with the system size, $k(L) \sim L^\alpha$.
      For $u = 1$ we have obviously $\alpha = 2/3$.

It is interesting to consider, again for $u = 1$, the density of contact points $g(x,L)$
for several lattice sizes and notice that for the function
\beq \label{I9}
G(x/L) = L^{\frac{1}{3}} g(x,L),
\eeq
 one has a data collapse. 
As suggested  by the conformal invariance for the confined critical systems 
\cite{BUXU}
we make an {\it ansatz} about the
functional behaviour of $G(x/L)$ taking:
\beq \label{I10}
 g(x,L) = [\frac{L}{\pi}\sin(\pi x/L)]^{-1/3} C(\cos (\pi x/L)).
\eeq
In Fig.~\ref{Ifig6}(b) 
one shows the function $C(\cos(\pi x/L)$. It is practically a constant. If
one assume that $C$ is independent of $\pi x/L$, one can get its value by
integrating $g(x,L)$ (Eq.~\rf{I10}) and compare the result with the average
number of clusters given by \rf{I8}. One obtains:
\beq \label{I11} 
C = -\frac{\sqrt 3}{6\pi^{\frac{5}{6}}}\Gamma(-\frac{1}{6}) = 
0.753149...,
\eeq
in very good agreement with the data obtained from Monte Carlo simulations
 (see Fig.~\ref{Ifig6}(b)).

The motivations for the {\it ansatz} \rf{I10} comes from the following 
observation.
 If the density of contact points corresponds to a local operator in
conformal field theory, one expects \cite{BUXU},
\beq \label{I12}
 g(x, L) = C [\frac{L}{\pi}\sin(\pi x/L)]^{X},
\eeq
where $X = \Delta + \bar{\Delta}$  is the scaling dimension of the local
operator ($\Delta-\bar{\Delta}$  is the conformal spin). Comparing 
Eqs.~\rf{I10}-\rf{I12} 
one concludes that the density of contact points corresponds to a local
operator with scaling dimensions $X = 1/3$. This result is surprising. One
can show that the profile of an operator with spin vanishes \cite{BUXU}. This
implies $\Delta = \bar{\Delta}  = 1/6$. This value can't be obtained from the
possible values of $\Delta$ (see \rf{3.5}) of minimal models. It might be
obtained from $W(A_n)$ algebras \cite{flohr,bouwknegt}
 but it is not
clear why W-symmetry should play any role in our problem. Another
possible explanation is that for a conformal theory with $c = 0$
(this is also the case of percolation), under certain circumstances, 
 the scaling dimension can be an
arbitrary number \cite{bauer1,bauer2}. 
 This possibility is a bit strange since
the number one has to obtain is a neat $1/6$. Finally, it is possible, that
the proof that operators with spin have no profile functions does not apply
to our problem. If this is the case, one can choose $\bar{\Delta}  = 0$ and
$\Delta = 1/3$ 
\footnote{For the appearance of
operators with conformal spin in the presence of the quantum group symmetry
mentioned in Sec. 3, see Ref.~\cite{AH}.}. 
 The value $1/3$ is a perfectly acceptable one (see Eq.~\rf{3.8} in Sec. 3).
To conclude, we have not yet an explanation for the value $X = 1/3$ in
 Eq.~\rf{I12}.

  We move away from $u =1$ and consider the domain $1 > w$. 
   The number of clusters $k(L)$ keep having an algebraic increase with an
  exponent $\alpha$ that  decreases monotonically with  $w$.
In Table~\ref{Itable1}   we give estimates for the exponent $\alpha$ for several values of
  $w$. It was  also observed that for several values of $w$  the scaling law
\beq \label{I13}
G(x/L) = L^{1-\alpha} g(x,L) 
\eeq
  holds.
\begin{table} 
\centerline{
\begin{tabular}{|c|c|c|c|c|c|c|c|c|}
  \hline
  w & 1 & 0.85 & 0.75 & 0.4 & 0.25 & 0.1 & 0.05 & 0.025\\\hline
  $\alpha$ & 0.67 & 0.50 & 0.40 & 0.30 & 0.24 & 0.15 & 0.06 & 0.01 \\
  \hline
\end{tabular}}
 \caption{
   Estimates of the exponent $\alpha$ giving the increase of the number
  of clusters with the size of the system $L$.
These data were obtained from numerical analysed of several lattice sizes 
up to $L= 65536$.}
\label{Itable1}
\end{table}

These results indicate that for $u>1$ 
the system stays scale invariant with a varying critical
   exponent.

\begin{figure}[t]
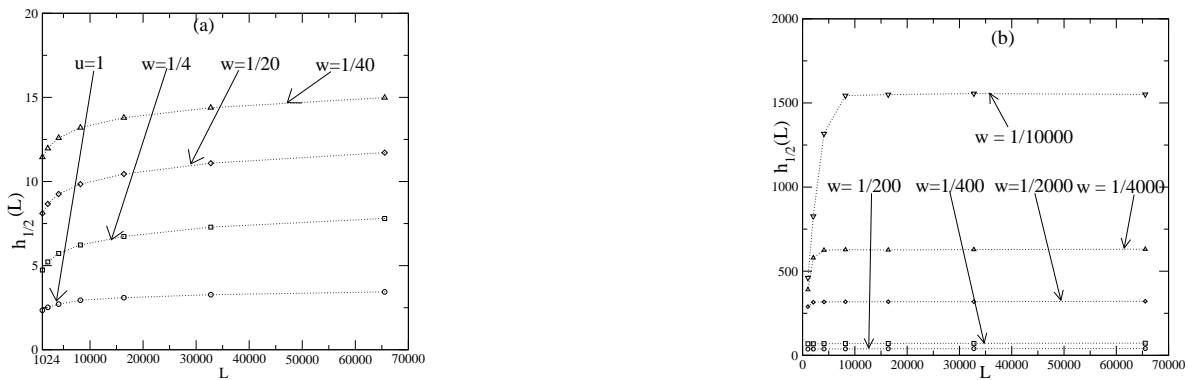

\centering{\includegraphics [angle=0,scale=0.35]
{sav19.eps}\hfill\includegraphics [angle=0,scale=0.35] {sav20.eps}}
\caption{
      (a) Average height in the middle of the lattice
     $h_{\frac{1}{2}}
     (L)$ as a function of the lattice size  $L$ for some  values
     of $w$ in the range
     $1 \geq w\geq 0.025$.
     (b)  Average height in the middle of the lattice
     $h_{\frac{1}{2}} (L)$
     as a function of the lattice size $L$ for various small values of $w$
     ($w \leq 1/200$).}
\label{Ifig10}\label{Ifig11}
 \end{figure}

    Although it seems obvious that one should look at the $L$ dependence of the
    average value of the heights $h(L)$ or the average value of the height at
    the middle of the lattice $h_{\frac{1}{2}}(L)$, it turns out that
    these quantities are
    harder to study since they vary slowly with the system size
(see \cite{ALR} for some illustrating figures).
In the case of $h_{\frac{1}{2}}(L)$ we have the dependence shown in Fig.~\ref{Ifig10}(a). 

\begin{figure}[th]
\centering{\includegraphics [angle=0,scale=0.49] {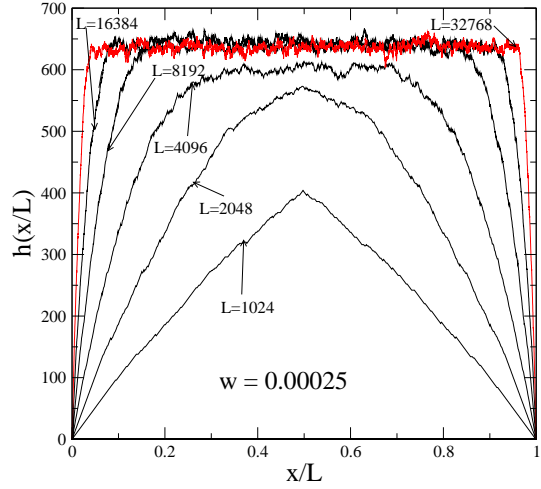}}
\caption{
(Color online) Typical configurations in the stationary state for $w = 0.00025$ and
different values of $L$.}
\label{Ifig12pp}
 \end{figure}

     A  different picture emerges if one considers very small values of $w$ as
     shown in Fig.~\ref{Ifig11}(b). One can observe two different phenomena. The values of
     $h_{\frac{1}{2}} (L)$ increase substantially
     as compared to the values observed at larger values of $w$,
     and more interestingly,
     it looks like
     they saturate for large values of $L$.
 We believe \cite{ALR} that this saturation phenomena happens in the whole region 
$u>1$ ($w<1$). We did not see such saturation for smaller  values of 
$u$ (see Fig.~\ref{Ifig10}(a)) because the lattice sizes considered in the 
simulations  was not large enough for these values of $u$. 
Considering these simulations with lattice size $L=65536$ a good fit, for various 
values of $w$, is obtained by the curve \cite{ALR}
\beq \label{I14}
 h_{\mbox{\scriptsize{max}}} = \frac{4.67}{ w^{0.62}}.
\eeq
  Interestingly,
  if we use (\ref{I14}) for $w=1$ we obtain $h_{\mbox{\scriptsize{max}}} = 4.67$, a
 value compatible with
   the results shown in Fig.~\ref{Ifig10}(a).
  Taking the results shown in Fig.~\ref{Ifig11}(b) at face value, implies that the
  phase SOC (see Fig.~\ref{afig4}) extends at least to the value
  $w = 10^{-4}$. On the
  other hand for $w=0$ the single configuration is the full triangle,
  $h_{\mbox{\scriptsize{max}}}$ has values of the size of the system and therefore the
  $L$-dependence of  $h_{\mbox{\scriptsize{max}}}$
  shows no saturation.
  This suggests a non-analytical behaviour of various quantities describing the system
  at $w = 0$. In order to illustrate this point, in Fig.~\ref{Ifig12pp}
  we show typical configurations
  in the stationary state for a very small value of $w$ ($w = 0.00025$) and
  different system
  sizes. One can see a
  clear change in the role of the boundaries if $L$ varies from 1024 to 32768.
  For $L = 1024$
  one is close to the full triangle configuration (expected for $w = 0$).
  However, as one increases
  $L$, one gets a plateau (in which the height is almost constant and $L$
  independent) and the
  boundaries play a less and less important role.

\section{ Space-time phenomena and avalanches }
\label{sec4}
 In the last section we present results that corroborate the phase diagram shown in Fig.~\ref{afig4}. 
In the present section we give additional facts supporting these 
results.
  We are going to use the Family-Vicsek \cite{FAVI}
  scaling in order to check if
  the system is in a scale invariant phase and if the answer is
  positive, determine the dynamical critical exponent $z$.
   If we consider a time-dependent average quantity $a(t,L)$, being 
     $a(L)$ the average in the stationary state, than for large
   values of $t$ and $L$ the function $A(t,L)$ scales as:
\beq \label{I15}
A(t,L) = \frac{a(t,L)}{a(L)} -1 \sim A(\frac{t}{L^z}),
\eeq
   where $z$ is the dynamical critical exponent. The scaling function depends on
   the initial conditions. In the data shown below, we have taken as initial
   condition the substrate ($h(2i) =0, h(2i+1)=1$, ($i=0,\ldots,n-1$), $h_L=0$)
   with probability one.
    We first consider the quantity $K(t,L)$ which corresponds to the
    average number of clusters ($a(t,L)$ in (\ref{I15}) is $k(t,L)$ in this case).

\begin{figure}[t]
\centering{\includegraphics [angle=0,scale=0.33]
{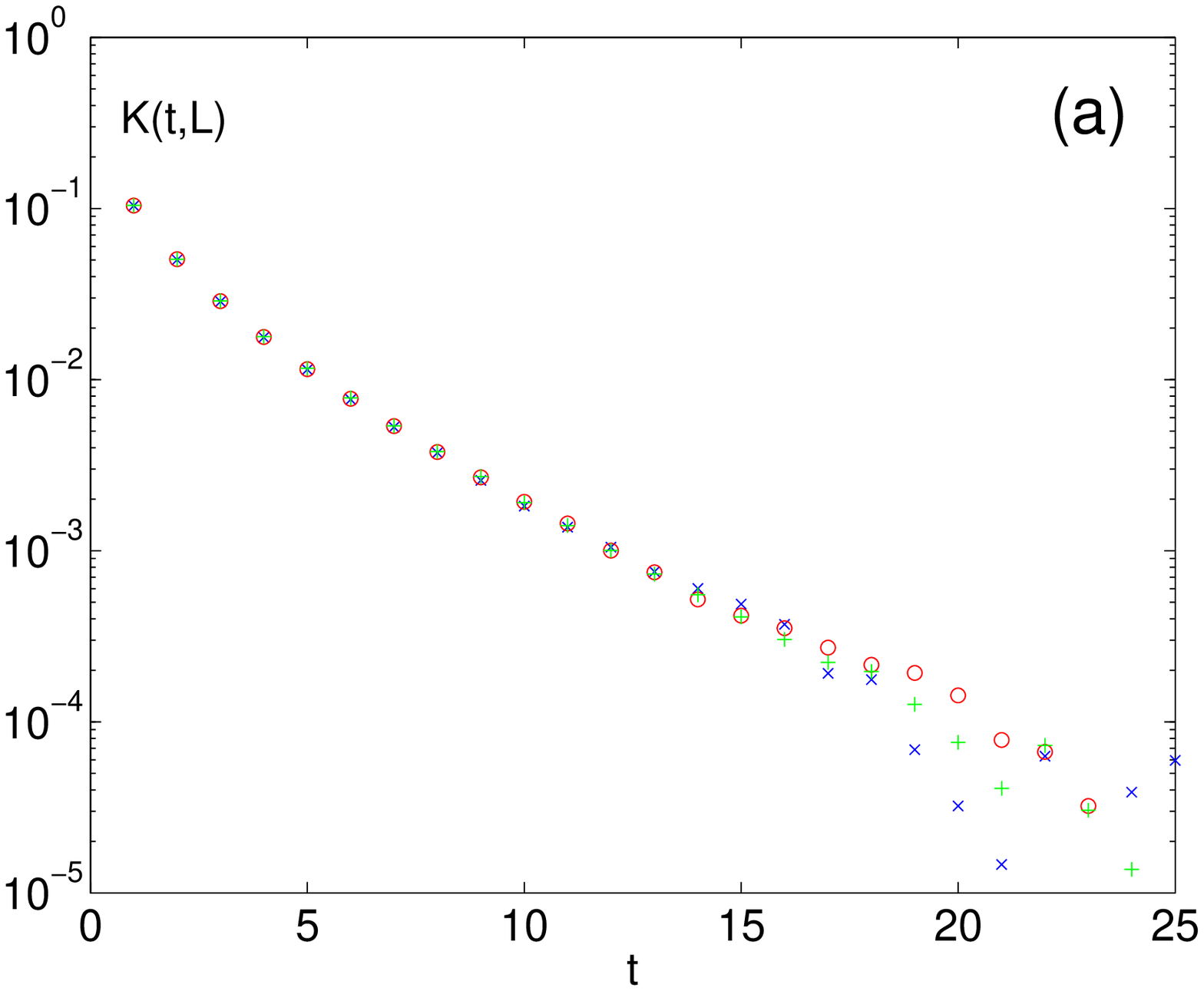}\hfill\includegraphics [angle=0,scale=0.44]{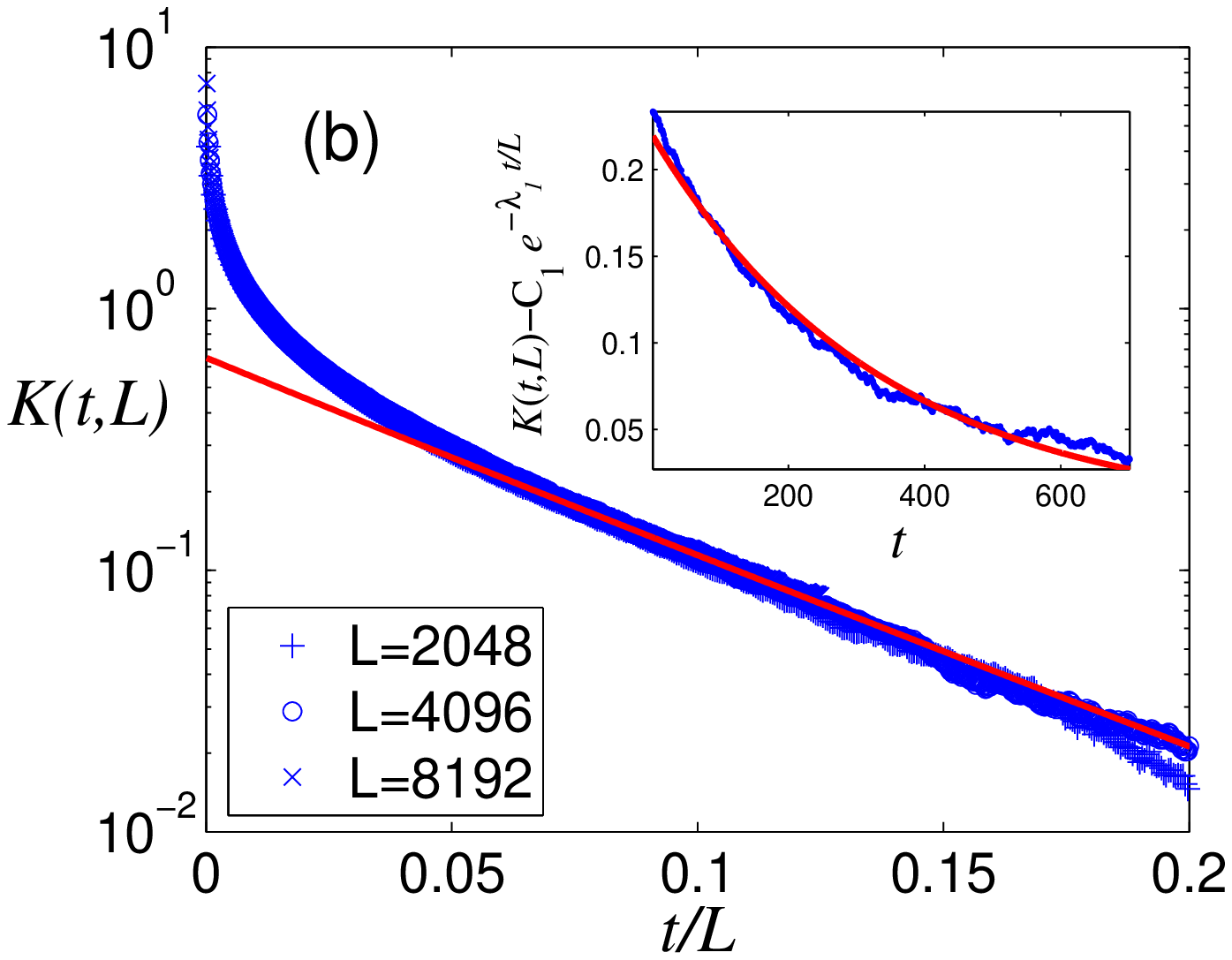}}
\caption{ (Color online) (a)     Function $K(t,L)$, defined on (\ref{I15}),
 for $u=0.4$ as a function of time for
    the lattice sizes $L=$ 2048, 4096 and 8192. 
    (b) Function $K(t,L)$ for $u=1$ as a
      function of $t/L$ for various lattice sizes. Red curve is given by a 
fit to (\ref{I15}) considering only the first exponential  
with $\lambda_1=\pi 3\sqrt{3}$ and $C_1 = 0.18$. In the inset 
 we compare $K(t,l)-C_1 e^{-\lambda_1 t/L}$ for $L=8192$ with the short time exponential $C_2 e^{-\lambda_2 t/L}$, given in red line (color online) with 
$\lambda_2= \frac{3}{2}\lambda_1$ and $C_2 = 0.54$.  }
\label{Ifig12}\label{Ifig13}\label{Ifig14}
 \end{figure}

 In   Fig.~\ref{Ifig12}(a) we show the function $K(t,L)$ for $u = 0.4$ for different lattice
    sizes.
%    ________________________________________________________________________
%    ______________________________________________________________________
We notice that for large lattice sizes, $K(t,L)$ is a function on $t$ only
     and can be obviously fitted with a sum of exponential functions as
     expected in a massive phase.
     Fig.~\ref{Ifig13}(b) shows $K(t,L)$ for $u = 1$ and as  expected from conformal
      invariance, $z=1$.
Conformal invariance gives also information on the function $K(t,L)$.
        One can use the knowledge of the finite-size scaling limit spectrum and obtain
       the functional dependence of $K(t/L)$.
      To illustrate how
      it works see Fig.~\ref{Ifig14}(b). For large values of t/L one can do a fit to the
      data, using the following ansatz:
\beq \label{I16}
K(t/L) = C_1 e^{-\frac{\lambda_1 t}{L}}
+  C_2 e^{-\frac{\lambda_2 t}{L}},
\eeq
      where the $\lambda_i$ are obtained from the spectrum of the  Hamiltonian (\ref{4.4}), given by \rf{3.7}:
\beq \label{I17}
E_i = \frac{\lambda_i}{L}, \; i=1,2.
\eeq
      The  constants $C_i$
      depend on the initial conditions. If one starts with the substrate 
      $C_1 = 0.28$ and  $C_2 =
      0.54$.

       We now consider smaller values of $w$ (larger values of $u$).
       We have done an
       extensive study of various functions $A(t,L)$ taking $K(t,L)$,
       $H(t,L)$ and
       $H_{\frac{1}{2}}(t,L)$ in  (\ref{I15}).
       The estimated values for the critical exponent $z$
       are shown in Table~\ref{Itable2}. For very small values of $w$, the estimates
       coming from $K(t,L)$ are poor because the number of
       clusters is small and hence
        large errors.
\begin{table}[t]
\centerline{
\begin{tabular}{|c|c|c|c|}
\hline
$w$ & $K$ & $H$ & $H_{1/2}$\\
\hline
1 & 1.03 & 0.95 & 0.97  \\
0.9 & 0.86 & 0.74 & 0.72 \\
0.7 & 0.55 & 0.52 & 0.50 \\
0.4 & 0.35 & 0.34 & 0.34 \\
0.25 & 0.31 & 0.30 & 0.30 \\
0.025 & * & 0.07 & 0.07 \\
\hline
\end{tabular}}
\caption{
       Estimates of the dynamical critical exponent $z$ 
for various values of $w$. The
       estimates were obtained using  (\ref{I15}) with $A$ replaced by  $K$, $H$ and $H_{\frac{1}{2}}$.   The number represented by (*) is not reliable since in that region the
       number of clusters is very small. 
These estimates were obtained by considering several lattice sizes up to $L=65536$.}
\label{Itable2}
\end{table}
Inspection of table \ref{Itable2} shows a smooth drop of the values of
    $z$  from $z=1$ for $w=1$ to $z \approx 0$    for $w \approx 0$. 
The
    latter values is in agreement with a direct calculation of the
    mass gap at $w=0$ \cite{GNPR1} which gives a finite gap in the
    thermodynamic limit. 
%insertion of the first referee observation
In the region where $z \to 0$ the information propagates, through the 
system, at infinite velocity. This can be explained nicely by Fig.~\ref{Ifig12pp} 
for large $L$. If the dynamics chooses a site near the edges, the peel process 
affects almost the whole interface, simultaneously lowering the entire 
upper plateau by a unit, implying that the information travelled an infinitely 
long distance in the limit $L \to \infty$ \cite{thankreferee}.

\begin{figure}[t]
\centering{\includegraphics [angle=0,scale=0.39] {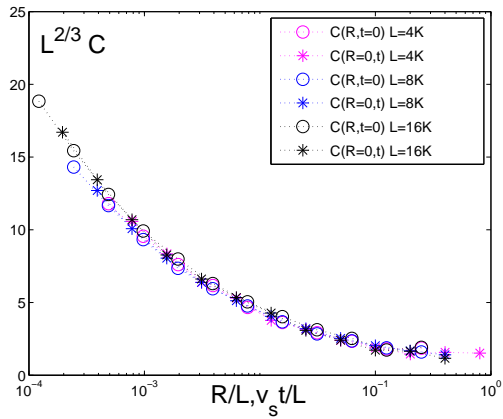}}
\caption{
       (Color online) The two-contact point correlation function
      $C(R,t,L)$
      times $L^{\frac{2}{3}}$ as a function of the scaled distance
      $R/L$ and the scaled time
      $\frac{v_s}{L}t$ ($v_s = \frac{3}{2}\sqrt{3}$ is the sound velocity)
      for various lattice sizes.}
\label{Ifig17}
 \end{figure}

     We have not studied extensively the two-contact points
 space-time correlation functions $c(R,t,L)$,
      except
      for the special case $u = 1$. This case is interesting since 
from conformal invariance it 
        exhibits, 
far from the  boundaries, 
       the following scaling form in the bulk (see (\ref{I9})):   
\beq \label{I18}
C(R,t,L) = L^{-\frac{2}{3}} G(\mu),
\eeq
      where
\beq \label{I19}
\mu = \frac{\sqrt{R^2 +v_s^2t^2}}{L}.
\eeq
      In (\ref{I18}) and (\ref{I19})  $R$ is the distance between the contact
      points, $t$ is the time difference and $v_s= \frac{3}{2}\sqrt{3}$ is
      the sound velocity \cite{ALR}.
      The factor $L^{-\frac{2}{3}}$ in (\ref{I18}) is obtained from  (\ref{I9}).
      In Fig.~\ref{Ifig17} we show the scaling function $G(\mu)$, where we
       have  used the
      unconnected correlation function in (\ref{I18}).
      To avoid boundary  effects the data was taken only in the center segment of size $L/2$.

%     _______________________________________________________________
%    _________________________________________________________________

       The functional dependence (\ref{I18}) which is special for the conformal
       invariant case is nicely seen in Fig.~\ref{Ifig17}.
       Our data are not good enough in
       order to extract from the small $\mu$ behavior of $G(\mu)$,
       the two-contact
       point correlation function in the thermodynamic limit.

 An interesting property of the RPM is the occurrence of avalanches in the
 desorption processes. For the case $u = 1$, based on exact results on
 small lattices, the existence of avalanches was  suggested in
 \cite{GNPR1}.
 The   Monte Carlo simulations on large
 lattices show  avalanches in the whole
  domain $0 < w \leq 1$.

  In the stationary state, once a tile from the rarefied gas hits the
  interface, it can be reflected, adsorbed or can trigger a nonlocal
  desorption process in which many tiles may leave the interface, this defines 
an avalanche. In the latter
  case the size of the avalanche is given by the number of tiles $T$ that 
are released in the process. 
  This number is
  always odd, therefore it is convenient to write: $T = 2v -1$
  ($v =1,2,\ldots$). For $u<1$ only a finite number of tiles are
  removed, since the density of clusters is finite. For $w\leq 1$, the
  cluster density vanishes, and therefore macroscopic number of
  tiles may be desorbed, hence  macroscopic avalanches.

   We denote by $S(v,L)$ the PDF which gives the probability for an
   avalanche
   of size $v$ for a system of size $L$. In models of self-organized
   criticality
   (SOC) \cite{BTW}-\cite{PIP}, for large values of $v$ and $L$,
   one may expect
   the PDF to exhibit a simple scaling law,
 \beq \label{I20}
 S(v,L) = v^{-\tau} F(\frac{v}{L^D}),
 \eeq
   characterized by two exponents $\tau$ and $D$.

\begin{figure}[t]
\centering{\includegraphics [angle=0,scale=0.45] {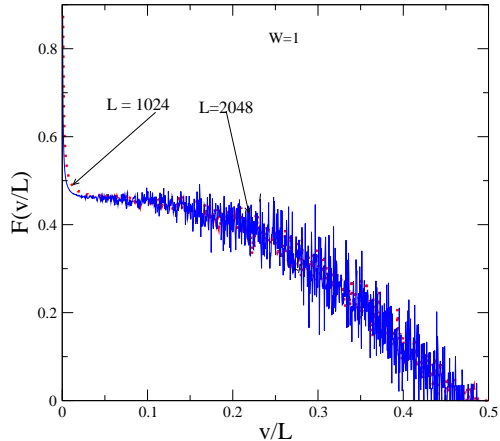}}
\caption{
(Color online) The scaling function $F(v/L)$ for $w=1$. The data are for lattice
sizes $L=1024$ and $L=2048$.}
\label{Ifig10-1}
 \end{figure}

The Monte Carlo simulations \cite{ALR} allow the evaluation of the above exponents.
For $u=1$ we have $\tau \approx 3$ and $D\approx 1$.
The result $D=1$ is to be
    expected since in a conformal invariant theory (this is the case if
    $u = 1$) one has no other length than the size of the system $L$.
    We found no
    explanation for the value three of $\tau$ although this number
    shows up in
    many aspects of the model (see for example  (\ref{I9})). 
    In Fig.~\ref{Ifig10-1} we show the scaling function $F(v/L)$
    defined in  (\ref{I20}). One observe a nice data collapse.

     In Table~\ref{Itable3} we give the estimates for the exponent $\tau$ and $D$
     for various
     values of $w$. The estimates presented in this Table are obtained using only
     one pair of sizes (4096 and 8192) but we made  sure that the
     results are
     reliable  by studying the changes of the values of the estimates
     using smaller lattices.
%_______________________________________________________________________________
%
\begin{table}[t]
\centerline{
\begin{tabular}{|c|c|c|c|c|c|c|c|c|c|c|}
  \hline
  $w$ & 1 & 0.95 & 0.90 & 0.80 & 0.60 & 0.45 & 0.30 & 0.05 & 0.02 & 0.005\\
  $D$ & 1.004 & 1.089 & 1.087 & 1.066 & 1.040 & 1.026 & 1.017 & 1.002 & 1.002 & 1.006 \\
  $\tau$ & 3.00 & 2.77 & 2.63 & 2.47 & 2.32 & 2.25 & 2.18 & 2.07 & 2.046 & 2.00 \\
  \hline
\end{tabular}}\caption{
      Estimates of the critical exponents $D$ and $\tau$ obtained
      by using lattices sizes
      $L = 4096$ and $L' = 8196$.}
\label{Itable3}
\end{table}

The results of Table~\ref{Itable3} indicate that the exponent $D$
 stays unchanged and equals to one in the whole 
range $w < 1$. This can be understood in the following way 
\cite{dhar-private}
 the tiles which are desorbed and create the avalanche 
belong always to a one-dimensional layer.

\begin{figure}[t]
\centering{\includegraphics [angle=0,scale=0.39] {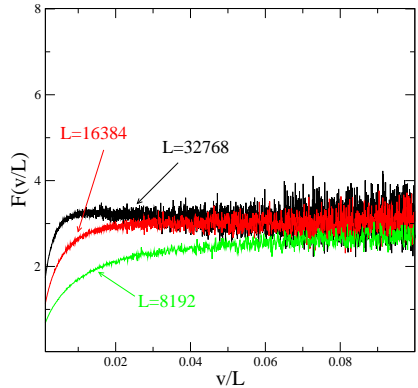}}
\caption{
 (Color online) The functions $F(v/L)$, defined on (\ref{I20}),
 at $w=0.00025$  for some values of $L$. }
\label{Ifig20p}
 \end{figure}

 We turn now to the exponent $\tau$. As can be seen in Table~\ref{Itable3} this exponent
varies from the value 3 ($u = 1$) to a value close to 2 for $w$
close to zero. In the case $u = 1$, the average size of the
avalanches stays finite when $L$ gets large but the dispersion
diverges (the function $F(v/L)$ shown in Fig.~\ref{Ifig10-1} does not
vanish at the origin). Does an exponent $\tau = 2$ mean that the
average size of the avalanches  diverges logarithmically with $L$ ?
Not necessarily, it depends if the scaling function $F(v/L)$
vanishes or not at the origin. In Fig.~\ref{Ifig20p} we show for $w =
0.00025$, that  $F(v/L)$ does not vanish at the origin. Since later in
this section we are going to show that $\langle v \rangle$ stays
finite for any $w >0$, we conclude that $\tau$ gets very close to
the value 2 but never reaches it.
For a more precise determination of the exponent $\tau$, see 
Figs. 15-18 in 
Ref.~\cite{ALR}. 

The Monte Carlo simulations show avalanches 
 at least 
in the interval $1 \geq w \geq 0.00025$ and therefore the system is scale 
invariant in this interval.
It is probably safe to assume 
that the 
SOC phase covers the domain $1 \geq w > 0$. 
It was   checked that for $u<1$, where the 
      system is massive, for large values of $v$ the PDF has an exponential
      decrease and not an algebraic one.

       We turn now our attention to the average size of the avalanches. A
       simple mean-field argument allows to compute it using informations
       about
       the stationary state only.

        We first consider, in the stationary state, the average fraction of
         interface where desorption does not take place $n(L)$ (FND)
         defined in
        (\ref{I6}). This quantity   varies from the
        value 1 for the substrate
        (the single
        configuration for $u = 0$) to the value $1/(L-1)$ for the full
        triangle (the
        single configuration for $w = 0$).

         For $u = 1$ there is a conjecture \cite{STRO} for the values
         of $n(L)$
\beq \label{I25}
n(L) = \frac{3L^2-2L+2}{(L-1)(4L+2)} =\frac{3}{4}(1-\frac{1}{6L}+\cdots ).
\eeq
          This conjecture was checked using our Monte Carlo simulations
          for
          various lattice sizes.

           If one knows $n(L)$, the average probability $P_a(L)$ to have
           an adsorption
           process, the average probability $P_r(L)$ to have a reflection
           process and
           the average probability $P_d(L)$ to have a desorption process
           can be easily
           obtained:
\beq \label{I26}
P_a(L) =  \frac{n(L)}{  2} - \frac{1}{2(L-1)},
\eeq
\beq \label{I27}
P_r(L) = \frac{n(L)}{2}(2w-1) +1-w + \frac{1}{2(L-1)},
\eeq
\beq \label{I28}
P_a(L) + P_r(L) + P_d(L) = 1,
\eeq
where we choose $u_a =1$ and
$u_d=w=1/u$. 

            Taking into account that in the stationary state the average
        number of
        adsorbed tiles is equal to the average number of desorbed
        tiles, one
        obtains:
\beq \label{I29}
\langle T\rangle_L = 2\langle v\rangle_L -1 = \frac{P_a(L)}{P_d(L)}.
\eeq
Note that this is a mean-field calculation since we have first computed 
the average probability to have a desorption process and multiplied it 
with the average number of tiles which are desorbed.
       In the large $L$ limit (\ref{I26})-(\ref{I29}) gives:
\beq \label{I30}
\langle T \rangle_{\infty} \approx \frac{n_{\infty}}{2(1-n_{\infty})w}.
\eeq

\begin{figure}[h]
\centering{\includegraphics [angle=0,scale=0.39] {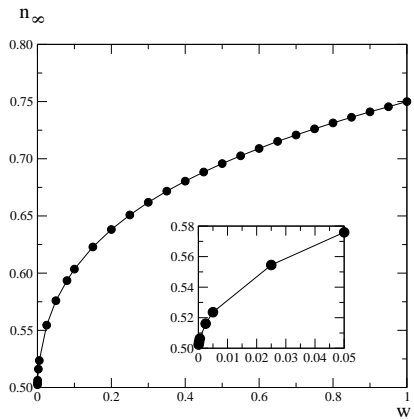}}
\caption{
   The $w$ dependence of the fraction of the interface where
 adsorption does not take place $n_{\infty}$ obtained from
the extrapolated results of Monte Carlo
 simulations. }
\label{Ifig21}
 \end{figure}

          In Fig.~\ref{Ifig21} one shows the $w$ dependence of
          $n_{\infty}$ as obtained from
          the extrapolated results of our Monte Carlo simulations.
One notices a discontinuous behavior of $n_\infty$ around $w=0$. At
$w=0$ 
where in the stationary state one has only one configuration, which is the 
full triangle (see Fig.~\ref{fignova}),
one has $n_\infty = 0$, while $ \lim_{w \to 0_+} n_{\infty} =  0.5$.
  Using (\ref{I30}) one concludes,
in agreement with previous observations, that for any finite value  $w$, the
average size of the avalanches is finite.
% insertion in the revised version
We consider that the discontinuity of $n_{\infty}$ at $w = 0$ shown in 
 Fig.~\ref{Ifig21} is 
one of the most interesting and unexpected property of the model. The 
value $n_{\infty}= 0.5$ observed for $w = 0_+$ 
corresponds to an interface in which 
on the average we  have an equal number of sites where one has local 
minima, maxima, positive or negative slopes. It is not clear to us why this 
interface
is special. The existence of this discontinuity also gives supplementary 
support for the existence of the SOC phase. Studying the finite-size 
scaling properties of the Hamiltonian spectra for small lattices and 
different values of $w$ in Ref.~\cite{GNPR1}
 it was shown that for $w<1$ one does not  have 
conformal invariance. On the other hand, for $w = 0$, we know \cite{GNPR1}
 that the 
system has an energy gap. In spite of all our determinations of critical 
exponents in the domain $0 < w < 1$ implying that the system is gapless, one 
could argue that one sees crossover effects and that one can have a 
massive phase in
the whole domain $0 \leq w < 1$. The discontinuity observed in the 
behaviour of $n_{\infty}$ assures us that this is not the case. 
% end of insertion for revised version

 In Table~\ref{Itable4}  we compare in two cases the mean-field predictions for
$\langle T\rangle_L$ given in
(\ref{I30}) with the values measured in Monte-Carlo simulations. The agreement is
excellent. In the same Table the values for $n(L)$ for several values of $L$ are also
given. One notices that for $u = 1$ and $L = 1024$ one  already  gets, within five digits,  the
asymptotic value 3/4. This is not the case for $w = 0.00025$, where 
the finite-size effects are larger. 
Our numerical analysis indicate that the asymptotic behavior 
of $n(L)$ can be described by
\beq \label{I31}
 n(L) = n_{\infty} ( 1 - A(w)/L + \cdots),
\eeq
where $A(1) = 1/6$ and $A(w)$ diverges if $w \rightarrow 0$ (data not shown).

 We conclude this section with the following observation, which is necessary for the consistency of
the present
model for avalanches. If one looks at the expression (\ref{I27}) of the average
(over configurations) probability that a tile hitting the interface is reflected,
one sees that for small values of $w$, the probability approaches the value one. This
implies that to obtain large avalanches (small $\tau$) one has to wait a long
time. This is to be expected in models of SOC.

%
%T444444444444444444444444444444444
\begin{table}[t]
\centerline{
\begin{tabular}{|c|c|c|c|c|c|}
\hline
 $w$ &$L$ & $n_L$ & $\langle T\rangle_L$ &MF&
 $ \langle T\rangle_L/\mbox{MF}$   \\
\hline
1 & 1024 & 0.749  & 1.497 & 1.497 & 0.999 \\
1 & 2048 & 0.749  & 1.498 & 1.501 & 0.998 \\
1 & 4096 & 0.749  & 1.499 & 1.500 & 0.998\\
1 & 8192 & 0.749  & 1.499 & 1.499 & 0.999 \\
\hline
\hline
0.00025 & 1024 & 0.199  & 496.999 & 496.119 & 1.001 \\
0.00025 & 2048 & 0.313  & 914.346 & 913.820 & 1.001\\
0.00025 & 4096 & 0.407  & 1374.713 & 1373.856 & 1.001 \\
0.00025 & 8192 & 0.455  & 1674.877 & 1674.550 & 1.000 \\
\hline
\end{tabular}}
\caption{
            Values of $n_{\infty}$ and  the average number of
            tiles obtained from Monte
            Carlo simulations $\langle T\rangle_L$ as compared
             with the mean-field
            results (MF). The values are given for several lattice
            sizes $L$ and two
            values of~$w$: $w=1$ and $w = 0.00025$.}
\label{Itable4}
\end{table}

\section{ Conclusions} \label{sect8}

 We have shown that the one parameter $u = 1/w$ dependent RPM has an 
unusual phase diagram (see Fig.~\ref{afig4}). At $u = 1$ (the Razumov-Stroganov 
point) the model has many facets. Firstly, the stationary state is 
related to a two-dimensional equilibrium system. The details of this 
observation are still at the level of a conjecture. The fact that the 
normalization factor of the stationary PDF is related to an enumeration 
problem of a two-dimensional system (fully packed loops or alternating 
sign matrices) was recently proved rigorously \cite{VP1}-\cite{ZF2}. The 
two-dimensional system has many interesting combinatorial properties 
which are reflected in the PDF of the stationary state. Also at $u = 1$ the 
Markov process has a space-time symmetry (conformal invariance) which does not
exist for $u \neq 1$. 
 
 How does the combinatorial facet of the model can be used to derive meaningfull 
 physical 
quantities? Using small system sizes, one can "guess" expressions which 
can be checked by Monte-Carlo for large systems. One example is the 
probability to have $k$ clusters for a system of size $L = 2n$ (see 
Eq.~\rf{I8}).  
Actually this expression is a solution of some bilinear recurrence 
relations (Pascal hexagon \cite{PP}). For another example, see Eq.~\rf{I26}.

 The conformal invariance facet of the model has other consequence. It 
fixes the value of the dynamical critical exponent: $z = 1$. It also fixes 
the finite-size scaling behavior of the density of contact points (see 
Eqs.~\rf{I11} and \rf{I12}), the form of the two-contact point 
correlation function  
(see Fig.~\ref{Ifig17}) and the Family-Vicsek scaling functions 
(see Eq.~\rf{I17} and 
Fig.~\ref{Ifig12}(b) ).

 The RPM has been extended in several directions. In one of them, one has 
kept the 
Hamiltonian 
 \rf{4.4} but enlarged the vector space on which it acts to a 
larger left ideal. This has allowed us \cite{AR}, keeping conformal invariance,
 to study the move of a random 
walker in an unquenched disordered system as well as the annihilation 
reaction $A + A \to \emptyset$ in the same medium. 
 The same study for other models, without   
  conformal invariance, can be found   in  \cite{KKGH} and \cite{HLM}.
 
 Some very interesting results have been obtained 
 by extending the 
TL semigroup algebra to the one-boundary TL semigroup algebra 
\cite{MS}-\cite{DNR}. The 
configuration space is different and biased toward the one end of the 
system where a new generator is placed. The model gets new combinatorial 
facets and conformal invariance is maintained.

 Before closing this paper, we would like to comment on the phase 
diagram of the RPM (see Fig.~\ref{afig4}). 
If we perturb the system and go away from $u = 1$, one obtains 
a massive phase if $u < 1$. This happens with many conformal invariant 
models. If however we take $u > 1$, the system stays scale invariant 
with varying critical exponents without being conformal invariant. This 
phenomenon is new
and it is a consequence of the nonlocality 
of the model \cite{rivacardy}. 
 From a phenomenological point of view we think that the 
observation that for any finite $u$ the average height stays finite for 
large lattice sizes is very unexpected (for unweighted RSOS paths it 
increase like $L^{1/2}$). 
Keeping in mind that for $w=0$, in the stationary state, there exists 
 only one 
configuration which is a full triangle (see Fig.~\ref{fignova}) one  
is tempted 
 to believe that configurations where on "feels" the triangle effects  
 should play an important role. This is not the case 
(see Fig.~\ref{Ifig12pp}).

\section*{Acknowledgments}
This work was  supported in part by  FAPESP and 
CNPq (Brazilian Agencies), by the Deutsche Forshungsgemeinschaf (Germany), and 
by  The Australian Research Council.

\appendix
\setcounter{section}{1}
\section*{Appendix. Link diagrams and RSOS (Dick) paths}

 We consider a one-dimensional lattice with $L + 1$ ($L = 2n$) sites.  The 
configuration space of the fluctuating interface is given by RSOS paths 
(better known in combinatorics as Dick paths). They are defined by taking 
nonnegative heights $h_i$ which obey the restricted solid-on-solid (RSOS) 
rules
\beq
h_{i+1} - h_i = \pm1,\qquad h_0 = h_L = 0,\qquad h_i \geq 0.
\label{A.1}
\eeq
See Fig.~\ref{fignova} for an 
example.

\begin{figure}[ht]
\centerline{
\begin{picture}(370,40)
\put(0,0){\epsfxsize=160pt\epsfbox{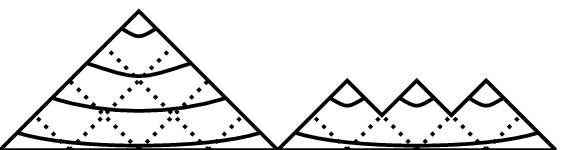}}
\put(170,0){$(a)$}
\put(210,0){\epsfxsize=160pt\epsfbox{conn.eps}}
\put(390,0){$(b)$}
\end{picture}}
\caption{ An interface with contour lines (a) and the corresponding
link pattern (b).}
\label{figA1}
\end{figure}

\begin{figure}[ht]
\centerline{
\begin{picture}(470,100)
\put(170,0){\epsfxsize=160pt\epsfbox{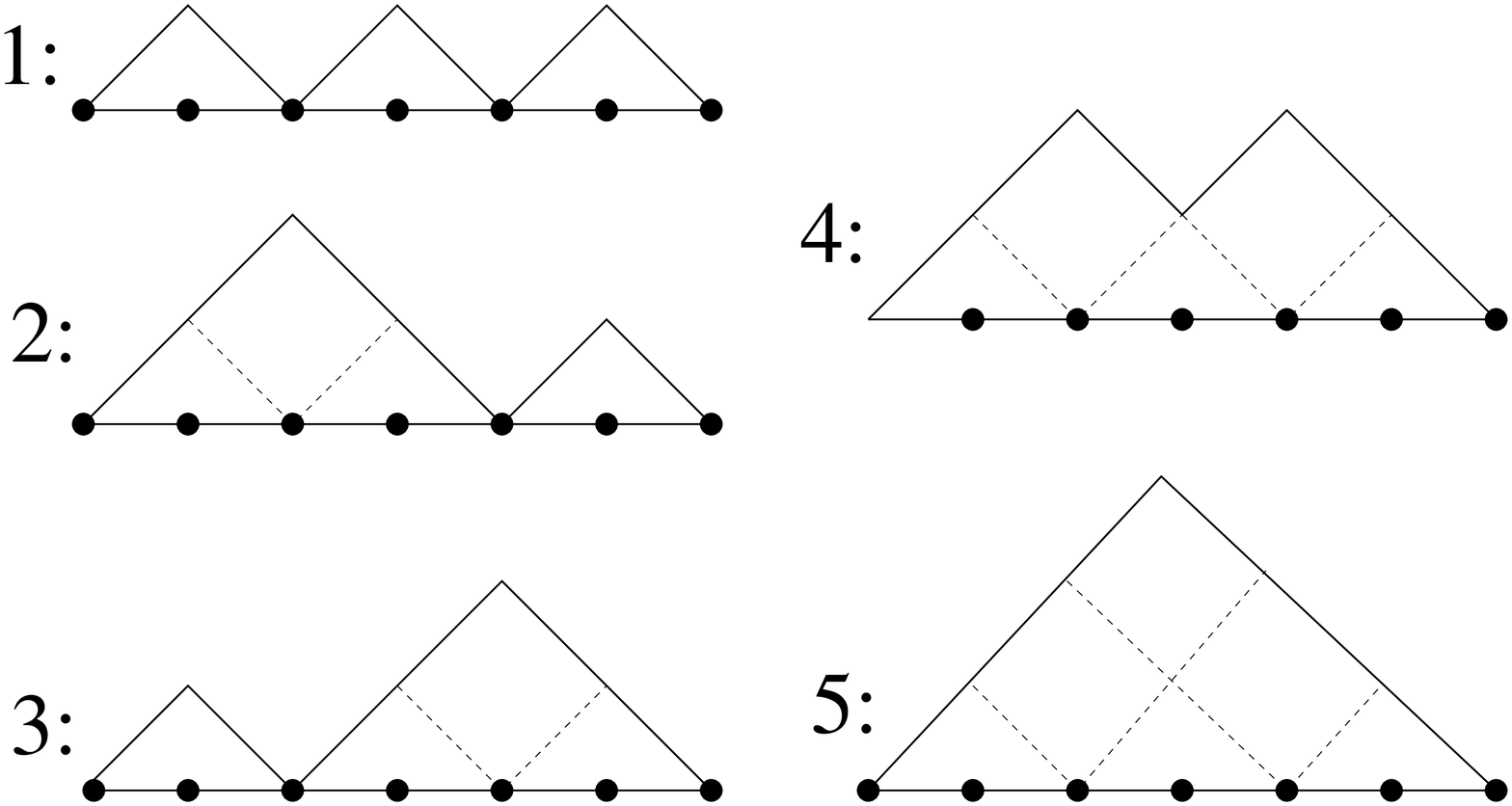}}
\end{picture}}
\caption{The RSOS configurations corresponding to the links patterns \rf{4.5} of 
the $L=6$ lattice ($(L+1)$ points).}
\label{fignova}
\end{figure}

 Each RSOS path corresponds to a link pattern (see section 3) in the 
following way. On each RSOS path, draw the equal height contour lines as 
in Fig. \ref{figA1}a. By straightening out the surface, keeping the contour lines 
and rotating the picture around the horizontal axis, we obtain Fig. 
\ref{figA1}b. 
The contour lines connect pairs of sites and Fig. \ref{figA1}b defines a link 
pattern. Notice than for an RSOS path defined on $L + 1$ sites corresponds a 
link pattern defined om $L$ sites.
%%%%%%%%%%%%%%%%%%%%%%%%%%%%%%%%%%%%%%%%%%%%%

 In the case $L = 6$ for example, to the five link patterns \rf{4.5} correspond
five RSOS configurations 

One can read off the action of the Hamiltonian \rf{4.6} on this 
configuration space and therefore get the different transition rates. 
These are the transition rates of the RPM at the Razumov-Stroganov point used
in Section 5.


\section*{References}
\begin{thebibliography}{99}


\bibitem{GNPR1}  de Gier J,  Nienhuis B,  Pearce P and  Rittenberg V, 2004 
{\it J. Stat. Phys.} {\bf 114} 1
\bibitem{ALR}  Alcaraz F C,  Levine E and  Rittenberg V, 2006 {\it J. Stat. Mech.} 
P08003
\bibitem{GNPR2} de Gier J,  Nienhuis B,  Pearce P A  and  Rittenberg V, 2003 
{\it  Phys. Rev.} E {\bf 67}
  016101
\bibitem{BGN}  Batchelor M T,  de Gier J and  Nienhuis B, 2001 
{\it J. Phys. A: Math. Gen} {\bf 34}
 L265
\bibitem{PRGN}  Pearce P A,  Rittenberg V,  de Gier J  and Nienhuis B, 2002  
 {\it J. Phys. A: Math. Gen.} {\bf 35}  L661 
\bibitem{EKLP}  Elkies N,  Kuperberg G,  Larsen M  and  Propp J, 1992 
{\it J. Algebraic Combin.} {\bf 1} 
\bibitem{BBNY} Batchelor M T ,  Bl\"ote H W,  Nienhuis B  and  Yung CM, 1996 
{\it J. Phys. A: Math. Gen.} {\bf 29} 
L399
  111;  219-234 
\bibitem{DG} de Gier J, 2005 {\it J. Discr. Math.} {\bf 298}  365
\bibitem{BR}  Bressoud D M,  1999 {\it Proofs and Confirmations: The story of the
Alternating Sign Matrix Conjecture} (Cambridge, Cambrige  Press)
\bibitem{RS1}  Razumov A V and  Stroganov Yu G, 2001 
 {\it  J. Phys. A: Math. Gen} {\bf 34}  3185
\bibitem{RS2}    Razumov A V  and   Stroganov Yu G, 2004 {\it  Theor. Math. Phys.} 
{\bf 138}  333 
\nonum Razumov A V  and   Stroganov Yu G, 2004 {\it Theor. Math. Phys.} {\bf 138}   395
\bibitem{RS3}  Razumov A V  and  Stroganov Yu G, 2005 {\it Theor. Math. Phys.}
 {\bf 142} 
 237 
\nonum Razumov A V  and  Stroganov Yu G, 
 2005 {\it Theor. Mat. Phys.} {\bf 142}  284  
\bibitem{CI} Di Francesco P,  Mathieu P, and  Senechal F, 1991 {\it Conformal 
Field Theory}, (Springer-Verlag: New York)
\bibitem{TL} Temperley H N V and  Lieb E, 1971 {\it Proc. R. Soc. London} A 
{\bf 322} 251
\bibitem{PM}  Martin P, 1991 {\it Potts models and related problems in statistical mechanics}, 
Series on Advances in Statistical Mechanics - v. 5, 
(Singapore: World Scientific)
\bibitem{DD} Dhar D, 1999 {\it Preprint}  cond-mat/9909009
\bibitem{MS} Martin P,  Saleur H, 1993 {\it Commun. Math. Phys.} {\bf 158} 155
\bibitem{PP}  Pyatov P, 2004 {\it J. Stat. Mech.}  P09003
\bibitem{DNR} de Gier J,  Nichols A  and  Rittenberg V, 2005 
 {\it  J. Stat. Mech.}  P03003
\bibitem{APR}  Alcaraz F C,  Pyatov P  and  Rittenberg V, in preparation
\bibitem{BJ}  Bish D and  Jones V, 1997 {\it Inv. Math.} {\bf 128} 89
\bibitem{PDF} Di Francesco P, 1998 {\it Nucl. Phys.} B {\bf 532} 609
\bibitem{BAT}   Batchelor M T,  de Gier J and  Nienhuis B, 2002 
\bibitem {MR}  Martins M J and  Ramos P B, 1997 {\it Nucl. Phys.} B {\bf 500}, 579
\bibitem{MNR}  Martins M J,  Nienhuis B  and  Rietman R, 1998 
{\it Phys. Rev. Lett.} 
{\bf 81} 504
\bibitem{dGN} de Gier J and   Nienhuis B, 2005 {\it J. Stat. Mech.}  P01006
\bibitem{PS}  Pasquier V and  Saleur H,  1990 {\it Nucl. Phys.} B {\bf 330} 523
\bibitem{AAA}  Alcaraz F C,  Barber M N,  Batchelor M T,  Baxter R J and 
 Quispel G R W, 1987 {\it J. Phys. A: Math. Gen.} {\bf 20} 6397 
\nonum 
 Alcaraz F C, 
 Barber M N and 
 Batchelor M T, 1988 {\it   Ann. Phys., NY}  {\bf 182} 
280 
\bibitem{BS} Bauer M  and  Saleur H, 1989 {\it Nucl. Phys.} B {\bf 320}  591
\bibitem{SIX}  Baxter R J, 1982 {\it Exactly solved models in statistical mechanics}, 
(Academic 
Press: London) 
\bibitem{KO}  Korepin V E, 1982 {\it Comm. Math. Phys.} {\bf 86} 391
{\it Int. J. Mod. Phys.}  B 
{\bf 16} 1883 
\bibitem{KU}  Kuperberg G, 2002 {\it Ann. of Math.}  {\bf 156} 835
\bibitem{BUXU}  Burkhardt T W and  Xue T, 1991 {\it Phys. Rev. Lett.} {\bf 66}
 895
\bibitem{flohr}  Flohr M, private communication
\bibitem{bouwknegt}  Bouwknegt P and  Schoutens K, 1993 {\it Phys. Rep.} {\bf 223}  183 
\bibitem{bauer1} Bauer M,  private communication 
\bibitem{bauer2}  Bauer M and  Bernard D,
2003 {\it Commun. Math. Phys.} {\bf 239}  493 
\bibitem{AH}  Arndt P F and  Heinzel T, 1995 {\it J. Phys. A: Math. Gen.} {\bf 28}  3567
\bibitem{FAVI}  Family F and  Vicsek E, 1985 {\it J. Phys. A: Math. Gen.} {\bf 18}  L75
\bibitem{thankreferee} We thank the referee of this paper for this nice observation
\bibitem{BTW}  Bak P,  Tang C and  Wiesenfeld K, 1988 {\it Phys. Rev.} A {\bf 38}   364
\bibitem{RRA}  Dhar D and Ramaswamy R, 1989 {\it Phys. Rev. Lett.} {\bf 63}  1659
\bibitem{DHA}  Dhar D, 1999 {\it Physica A} {\bf 263}  4
\bibitem{TMS}  Tebladi C, De Menech M and  Stella A L, 1999
{\it Phys. Rev. Lett.}
{\bf 83}  3952
\bibitem{PIP}  Priezzhev V B,  Ivashkevich E V,  Povolotsky A M  and  Hu C K,
 2001 {\it Phys. Rev. Lett.} {\bf 87}  084301
\bibitem{dhar-private} Dhar D, private communication
\bibitem{STRO}  Stroganov Yu G,  private communication
\bibitem{VP1} Pasquier V, 2006 {\it Ann. Henri Poincare} {\bf 7} 397
\bibitem{ZF1} Di Francesco P  and  Zinn-Justin P, 2005 {\it Electron.  J. Comb.} 
{\bf 12} R6 
\bibitem{ZF2} Di Francesco P and Zinn-Justin P, 2005 {\it J. Phys. A:  Math. Gen.}
 {\bf 38} L815 
\bibitem{AR} Alcaraz F C and Rittenberg V, 2006 {\it Preprint} cond-mat/0612272
\bibitem{KKGH} Kissinger T, Kotowicz A,  Kurz O,  
Ginelli F, and   Hinrichsen H, 2005 {\it J. Stat. Mech.}  P06002
\bibitem{HLM} Ginelli F,  Hinrichsen H,  Livi R,  Mukamel D,  Torcini A, 2006 
{\it J. Stat. Mech.}  P08008
\bibitem{rivacardy} Riva V and Cardy J, 2005 {\it Phys. Lett.} B {\bf 622} 339

\end{thebibliography}
\end{document}